\documentclass[12pt, a4paper]{article}
\usepackage[utf8]{inputenc}
\usepackage[T1]{fontenc}
\usepackage{lmodern}
\usepackage{amsmath, amssymb, amsthm}
\usepackage{graphicx}
\usepackage{booktabs}
\usepackage{threeparttable}
\usepackage{rotating}
\usepackage{float}
\usepackage{afterpage} 
\usepackage{caption}  
\usepackage[authoryear]{natbib}
\usepackage[colorlinks=true, citecolor=blue]{hyperref}
\usepackage{subcaption}

\makeatletter
\renewcommand{\@maketitle}{
  \newpage
  \null
  \vskip 2em
  \begin{center}
    {\normalfont\bfseries\Large\@title \par} 
    \vskip 1em
    {\normalfont\large\@author \par} 
  \end{center}
}
\makeatother

\usepackage[top=2.5cm, bottom=2.5cm, left=2.5cm, right=2.5cm]{geometry}
\usepackage{setspace}
\title{Rethinking Nonstationarity in Time Series: A Deterministic Trend Perspective}

\author{%
Zhandos Abdikhadir\textsuperscript{1*} and Terence Tai Leung Chong\textsuperscript{2†}%
}

\date{\today}

\begin{document}

\maketitle

\begin{center}
\small
\textsuperscript{1} Department of Mathematics, Hong Kong Baptist University, Hong Kong\\
\textsuperscript{2} Department of Economics, Chinese University of Hong Kong, Hong Kong
\end{center}

\begin{center}
\normalsize \today
\end{center}

\begingroup
\renewcommand\thefootnote{}
\footnotetext{* Email: \texttt{abdikhadirzhandos@gmail.com}}
\footnotetext{† Corresponding author. Email: \texttt{chong2064@cuhk.edu.hk}}
\endgroup

\begin{abstract}

This paper challenges the dominance of stochastic trend models by introducing the Seasonal–Trend–Stationary ARMA (STSA) framework, which represents univariate nonstationary time series as stationary fluctuations around deterministic trend and seasonal components, allowing for a finite number of structural breaks in the trend. We present methods for estimating the locations and number of breaks using a dynamic programming algorithm and a sequential prediction-interval-based procedure, respectively, and outline strategies for specifying and estimating the full model. Empirical analysis of U.S. exports of goods to Mainland China (2006–2025) demonstrates that the STSA model accurately identifies structural breaks linked to major economic events and provides a meaningful decomposition of the underlying economic cycle dynamics. Evaluation on the monthly M4 Competition data shows that STSA significantly outperforms Prophet and, while generally less accurate than stochastic trend models such as ARIMA, ETS, TBATS, and Theta, it produces superior forecasts for series with abrupt structural breaks where stochastic approaches struggle to adapt. Unlike traditional time series models, STSA offers an interpretable decomposition that reveals the causal narrative behind the series’ evolution, enhancing the credibility of out-of-sample forecasts.

\vspace{\baselineskip}

\end{abstract}

\thispagestyle{empty}
\newpage

\section{Introduction}

A common practice in time series analysis and forecasting is to model the trend component as a stochastic process rather than a deterministic function. This approach offers several advantages, including greater flexibility in capturing unpredictable movements and producing more reliable forecasts, as it does not constrain the series to follow a rigid functional form. The seminal study by \citet{nelson_plosser_1982} examined whether macroeconomic time series are better characterized as trend-stationary processes—exhibiting stationary fluctuations around a deterministic trend—or as unit root processes, displaying a stochastic trend that does not revert to a deterministic path. The key distinction between these two approaches lies in the treatment of shocks: the former assumes that historical random shocks are transitory, with their effects dissipating over time, while the latter treats shocks as having permanent effects on the system. Based on the unit root test proposed by \citet{dickey_fuller_1979}, their findings provided strong evidence that most macroeconomic time series are well characterized by unit root processes, indicating that it is more appropriate to allow the trend component to evolve over time rather than remaining fixed.

Despite the widespread acceptance of the unit root hypothesis, \citet{perron_1989} challenged this view by arguing that nonstationary series can often be regarded as trend-stationary once structural breaks are taken into account. Such breaks frequently arise from economic crises, policy shifts, technological disruptions, or pandemics that substantially alter the trajectory of the series. He argues that, instead of assuming all past shocks have permanent effects on the system, it may be more appropriate to assume that only a limited number of shocks are permanent. Consequently, instead of presuming that the trend always evolves or remains fixed, it is sometimes preferable to model the trend as changing only occasionally. \citet{perron_1989} treated the Great Crash of 1929 and the 1973 oil crisis as known structural shifts, allowing him to reject the unit root hypothesis for the majority of the macroeconomic series considered by \citet{nelson_plosser_1982}. He demonstrated that ignoring structural changes in the trend component reduces the power of traditional unit root tests and biases results toward falsely concluding the presence of a unit root when the series is, in fact, trend-stationary with breaks.

Subsequent work by \citet{zivot_andrews_1992}, \citet{banerjee_etal_1992}, and \citet{perron_vogelsang_1992} demonstrated that treating break dates as known a priori is inappropriate. They developed procedures to estimate the timing of structural shifts, revised the testing framework proposed by \citet{perron_1989}, and reported fewer rejections of the unit root null hypothesis compared to the initial findings. In contrast, \citet{lumsdaine_papell_1997} reignited the debate by finding stronger evidence against the unit root hypothesis when allowing for two breaks in the model. Since then, research has focused on developing more robust tests with improved size and power properties. The key insight of this debate is that models failing to adequately account for structural changes are misspecified and can therefore provide spurious evidence of strong persistence. This body of work demonstrates that proper treatment of structural breaks is crucial, as many series previously believed to follow a stochastic trend may instead follow a deterministic trend once sufficient structural shifts are incorporated.

Despite the large amount of theoretical research conducted on trend stationarity and changepoints, applications remain limited, underscoring the need for broader exploration of deterministic trend models as complements to the prevailing stochastic trend and unit root frameworks for nonstationary time series. A prime example of the latter approach is the Autoregressive Integrated Moving Average (ARIMA) model, popularized by \citet{box_jenkins_1970}, which remains one of the most widely used benchmark models in time series analysis and forecasting. To the best of our knowledge, no practical framework currently exists for modeling nonstationary time series as stationary fluctuations around the deterministic components.  Perhaps the closest approach to this idea is the Prophet model developed by Meta \citep{taylor_letham_2018}. However, the Prophet model does not explicitly assume that, after removing deterministic components, the remaining part is stationary, and its theoretical foundations and practical implementation have also been subject to considerable criticism within the time series community. Nevertheless, the model remains popular among practitioners, largely due to its interpretability.

This paper addresses this gap by introducing a practical framework that describes the nonstationary time series as stationary fluctuations around a deterministic trend and, if present, seasonality components, once an appropriate number of structural breaks in the trend are imposed. The proposed method provides a more interpretable insights than stochastic trend models, as it allows researchers to explicitely observe how structural changes affect the average growth rate of the series—a quantity of central interest. Linking the estimated breaks to events such as financial crises, policy reforms, or technological innovations provides a clearer causal narrative for understanding the evolution of the trend component. We discuss methods for estimating the locations of changepoints using a modified dynamic programming algorithm, as well as determining the number of breaks through a sequential prediction-interval-based algorithm. The proposed estimation procedures for the break parameters are validated through Monte Carlo simulations. Additionally, some guidance is offered for specifying and estimating the full model.

The empirical evaluation uses U.S. exports of goods to China (2002--2025) as an illustrative example to demonstrate how the proposed model selects the optimal configuration and decomposes the series into interpretable estimated components, capturing shifts in the underlying dynamics driven by geopolitical tensions, crises such as the COVID-19 pandemic and the global financial crisis, and other major events. We also evaluate the forecasting performance of the model by comparing it against established statistical approaches using monthly series from the M4 Competition \citep{makridakis_etal_2020}, highlighting cases in which the model generally outperforms or underperforms state-of-the-art methods. All methods and empirical analyses are implemented in Python.

The structure of the paper is as follows. Section~2 describes the model specification and proposed estimation procedures. Section~3 presents the empirical evaluation, including a detailed example of the model’s implementation and its forecasting performance. Finally, Section~4 provides discussion and concluding remarks.

\section{The Model and Estimators}

\subsection{Seasonal-Trend-Stationary ARMA}

We consider a classical additive structural model that decomposes a univariate nonstationary time series into three components: a deterministic trend component $\tau_t$, a deterministic seasonal component $s_t$, and the stochastic error term $u_t$. Formally, the model is written as
\begin{equation}
y_t = \tau_t + s_t + u_t
\label{eq:decomposition}
\end{equation}
In many applications, a multiplicative model may be more appropriate, which can be converted to an additive form \eqref{eq:decomposition} by applying a logarithmic transformation to the original series. We assume that the series exhibits stationary fluctuations around its deterministic trend component and, if present, its deterministic seasonal component, and thus describe it under the seasonal-trend-stationary hypothesis.

Since the underlying trend component is typically unknown, we model it as a piecewise linear function to parsimoniously capture long-term movements and occasional structural shifts. To evaluate the effect of each season, the seasonal component is represented by a set of dummy variables to account for recurring periodic patterns. The error term is assumed to follow a zero-mean, weakly stationary, and invertible ARMA($p,q$) process to capture the temporal dependence in the series. Together, the seasonal and error components describe the overall oscillations around the trend, with the former capturing the deterministic part and the latter capturing the stochastic part of the variation. We further assume that the seasonal and error parameters are constant, as structural shifts in these components are rare in practice and generally negligible relative to those in the trend component. Consider a seasonal time series $\{y_t\}_{t=1}^T$ with period $P$, featuring $m$ structural breaks occurring at times $(T_1, \dots, T_m)$, which divide the series into $m+1$ regimes. The components can then be expressed as
\begin{equation}
\tau_{t}=\begin{cases}
\mu_1+\beta_{1}t, & T_{0}<t\leq T_{1}, \\
\mu_{2}+\beta_{2}t, & T_{1}<t\leq T_{2}, \\
\vdots & \vdots \\
\mu_{m+1}+\beta_{m+1}t, & T_{m}<t\leq T_{m+1},
\end{cases}
\label{eq:trend}
\end{equation}
\begin{equation}
s_t=\sum_{j=1}^P \mathbf{1}\{s(t)=j\}\,\delta_j,
\label{eq:seasonality}
\end{equation}
\begin{equation}
\phi(L)u_t = \theta(L) \epsilon_t
\label{eq:error structure}
\end{equation}
Here, for $i = 1, \ldots, m+1$, $\mu_i$ and $\beta_{i}$ denote the intercept and the slope of the $i$-th regime, which starts at $t = T_{i-1}$ and ends at $t = T_i$ (with $T_0 = 0$ and $T_{m+1} = T$ for convenience). The breaks are assumed to take place at discrete time points. For $j = 1, \ldots, P$, the seasonal index function $s(t)$ assigns each time point $t$ to its corresponding season $j$, and $\mathbf{1}\{s(t)=j\}$ is an indicator variable equal to 1 if $s(t)=j$ and 0 otherwise, with $\delta_j$ representing the effect of season $j$. The polynomials $\phi(L) = 1 - \phi_1 L - \dots - \phi_p L^p$ and $\theta(L) = 1 + \theta_1 L + \dots + \theta_q L^q$ denote the autoregressive and moving-average operators of orders $p$ and $q$, respectively, with parameter vectors $\phi = (\phi_1, \dots, \phi_p)'$ and $\theta = (\theta_1, \dots, \theta_q)'$. All roots of $\phi(z)$ and $\theta(z)$ lie outside the unit circle to ensure weak stationarity and invertibility. Finally, $\{\epsilon_t\}_{t=1}^T$ is a white noise process satisfying $\mathbb{E}(\epsilon_t) = 0$, $\mathrm{Var}(\epsilon_t) = \sigma_\epsilon^2 < \infty$, and $\mathrm{Cov}(\epsilon_t, \epsilon_s) = 0$ for $t \neq s$.

We impose a continuity constraint on the trend component in \eqref{eq:trend} to ensure smooth transitions between adjacent regimes. This improves interpretability by removing abrupt level shifts and allowing a clearer assessment of trend evolution across regimes. The constraint also reduces the number of parameters by expressing each regime’s intercept (except for the first) in terms of the initial intercept, the break locations, and the slopes of the current and preceding regimes. Specifically, for \( i = 1, 2, \ldots, m\),
\begin{equation}
\mu_{i} + \beta_{i} T_i = \mu_{i+1} + \beta_{i+1} T_i
\label{eq:cont_constraint}
\end{equation}
Additionally, we impose a constraint on the seasonal coefficients in \eqref{eq:seasonality} to ensure they are interpreted as deviations from the average level:
\begin{equation}
\sum_{j=1}^{P} \delta_j = 0
\label{eq:seas_constraint}
\end{equation}

After imposing constraints \eqref{eq:cont_constraint} and \eqref{eq:seas_constraint}, Equation \eqref{eq:decomposition} can be expressed in the matrix form as
\begin{equation}
Y = X_{T^{(m)}} \beta + Z \delta + U,
\label{eq:multiple linear regression}
\end{equation}
where $Y = (y_1, \dots, y_T)'$, $U = (u_1, \dots, u_T)'$, and $T^{(m)}=(T_1, \dots, T_m)$. The trend component is represented by the parameter vector $\beta = (\mu_{1}, \beta_{1}, \ldots, \beta_{m+1})'$ and the matrix $X_{T^{(m)}}=(x_1, \dots, x_T)'$, where $x_t = (1, D_{t,1}, \dots, D_{t,m+1})$ and, for $i = 1, \dots, m+1$, $D_{t,i}$ is an indicator variable defined as follows:
\begin{equation*}
D_{t,i} =
\begin{cases}
0, & t \leq T_{i-1}, \\
t - T_{i-1}, & T_{i-1} < t \leq T_i, \\
T_i - T_{i-1}, & t > T_i
\end{cases}
\label{eq:vector_x}
\end{equation*}
The seasonal component is represented by the parameter vector $\delta = (\delta_1, \ldots, \delta_{P-1})'$ and the corresponding matrix $Z=(z_1, \dots, z_T)'$ such that $z_t$ is defined as
\begin{equation*}
z_t =
\begin{cases}
e_{s(t)}', & \text{if } s(t) \in \{1, \dots, P-1\},\\[1mm]
(-1, -1, \dots, -1), & \text{if } s(t) = P,
\end{cases}
\label{eq:vector_z}
\end{equation*}
where $e_{s(t)}$ is a standard basis vector.

The primary objective of this modeling approach is to obtain interpretable parameters that offer insights into the dynamics of the underlying series. Specifically, we aim to identify the timing of structural breaks and the associated magnitude of changes in the trend component, quantify the effects of seasonal patterns, and characterize the temporal autocorrelation structure among successive observations. If the number of breaks $m$ and their time indices ($T_1, \ldots, T_m$) are known, the model reduces to a classical multiple regression with ARMA errors. In practice, however, the break parameters are rarely known and must also be estimated. Unfortunately, simultaneous estimation of all parameters is computationally intensive, primarily due to the discrete nature of break locations, which causes the objective function of traditional methods such as least squares (LS) or maximum likelihood (ML) to be non-smooth, making gradient-based optimization problematic. To address this challenge, we adopt a three-stage procedure. First, we estimate the break locations across multiple candidate models. Next, we determine the optimal number of breaks, and finally, we provide efficient estimates of the remaining model parameters conditional on the chosen partition. We demonstrate that this procedure yields reasonable estimates of the true model parameters.

\subsection{Estimating Break Locations}

Research on structural breaks dates back to \citet{quandt1958, quandt1960} and \citet{chow1960}, who developed formal procedures for estimating and testing a single structural change in regression models. \citet{gallant_fuller_1973} proposed a method for determining break locations in segmented polynomial regression models using LS estimation, and \citet{feder1975} extended this line of work by developing the corresponding asymptotic theory. \citet{bai_perron_1998, bai_perron_2003} further provided a comprehensive framework and computationally efficient algorithms for estimating and testing multiple structural breaks in stationary regression models. For a comprehensive review of the theoretical development on structural breaks, see \citet{perron_2006}, and for an overview of recent advances in break detection methods, see \citet{truong2020}.

We consider a method for estimating break locations based on the global LS criterion. If the number of breaks $m$ is known, the estimates $\hat{T}^{(m)} = (\hat{T}_1, \dots, \hat{T}_m)$ are defined as
\begin{equation}
\hat{T}^{(m)} = \arg\min_{T^{(m)}} SSR(T^{(m)}),
\label{eq:break estimator}
\end{equation}
where $SSR(T^{(m)}) = Y'(1-P_{T^{(m)}})Y$ is the sum of squared residuals for a given partition $T^{(m)}=(T_1, \dots, T_m)$, and  $P_{T^{(m)}}$ is the corresponding projection matrix, obtained by ordinary least squares (OLS), defined as $P_{T^{(m)}} = A_{T^{(m)}}(A_{T^{(m)}}'A_{T^{(m)}})^{-1} A_{T^{(m)}}'$ with $A_{T^{(m)}} = [X_{T^{(m)}} \; Z]$. The properties of the LS estimators are formulated in terms of the estimated break fractions, $\hat{\lambda}_{i} = \hat{T}_{i} / T$, for $i = 1, 2, \ldots, m$. The asymptotic framework assumes that as the sample size $T$ increases, all segment lengths grow proportionally, preserving their relative sizes. \citet{bai_perron_1998} demonstrated that, under general assumptions on regressors and break fractions---primarily concerning their identifiability---the global LS estimators of the break fractions are super-consistent. Specifically,
\begin{equation}
T(\hat{\lambda}_{i} - \lambda_{i}^0) = O_p(1),
\label{eq:break_consistency}
\end{equation}
where $\lambda_{i}^0 = T_{i}^0 / T$ denotes the true break fraction of the $i$-th break, and $T_{i}^0$ is its true location. It is important to note that the $T$-rate convergence applies to the estimated break fractions $\hat{\lambda}_{i}$, not to the break locations $\hat{T}_{i}$. For the latter, the difference $|\hat{T}_{i} - T_{i}^0|$ is bounded by a constant independent of the sample size, with a probability arbitrarily close to one in large samples.

Although originally derived for stationary regression models, these results can be extended to the case of trending regressors through appropriate scaling to ensure bounded trends. \citet{perron_qu_2006} further showed that introducing linear constraints on the parameters does not violate the regularity conditions or affect the consistency of the global LS estimates. This result is crucial for our case, where we impose a continuity constraint on the trend and restrict the seasonal parameters to be stable across all segments. For the case of a piecewise continuous unbounded trend, \citet{perron_zhu_2005} showed that the LS estimators converge at the faster rate of $T^{3/2}$, although their result was derived for the single-break case. Notably, \citet{deng_perron_2006} demonstrated that in the continuous-trend case, the asymptotic properties of the break fraction remain the same for both bounded and unbounded trend specifications. Although, as far as we are aware, a formal derivation establishing the consistency of multiple break fraction LS estimators under a continuous-trend constraint has not been developed, we assume a convergence rate of at least $T$, which is a conservative yet reasonable assumption.

Solving the optimization problem \eqref{eq:break estimator} directly through a grid search of all possible partitions is computationally demanding, as it involves least-squares operations of order \(O(T^m)\). To overcome this limitation, \citet{bai_perron_2003} discuss an efficient dynamic programming (DP) algorithm that substantially reduces computational complexity, particularly when \(m > 2\). The method finds the global minimizers by exploiting the additive structure of the objective function and the independence between segments. It works recursively, using the solutions to smaller subproblems to efficiently solve larger ones, requiring at most \( O(T^2) \) least-squares operations for any number of structural changes $m$. The algorithm was originally developed for pure structural models, where all parameters are subject to shifts without constraints. Therefore, the DP algorithm cannot be applied directly in our setting, as the continuity constraint on the trend parameters and the assumption of fixed seasonality parameters introduce interdependence across segments. In this scenario, a full grid search is necessary to obtain the global minimizers, as all regimes are mutually dependent.

We propose a modified version of the DP algorithm that partially accounts for the parameter constraints while preserving the computational efficiency of the original method. The key idea is to store the estimated breaks within subsamples and recompute the objective function to adjust for dependencies across segments. The optimization of the objective function \eqref{eq:break estimator} is taken over the set of admissible partitions \(T^{(m)} = (T_1, \ldots, T_m)\) such that, for \(i = 2, \ldots, m\), \(T_i - T_{i-1} \geq h_1\), where \(h_1\) denotes the minimum allowable length of a single segment. This constraint is imposed to prevent the estimation of breaks that are too close to each other. In addition, we further restrict the search for the first break to points satisfying \(T_1 \geq l_1\) and for the last break to points satisfying \(T_m \leq T - l_2\), where \(l_1, l_2 \geq h_1\); these constraints help determine the appropriate number of breaks and produce more reliable forecasts, as will be discussed in more detail in the following sections. Generally, the parameter $h_1$ should not be too large, as this can lead to inaccurate estimates of break locations. Furthermore, the parameters $l_1$ and $l_2$ should be chosen carefully, as they may prevent the detection of breaks near the start and end of the sample, respectively. These parameters can be chosen visually by inspecting a plot of the series.

As an example, consider a series of length \(T = 100\) that contains two structural changes, with search constraints \(h_1 = 5\) and \(l_1 = l_2 = 10\), respectively. In the first step, we consider single-break partitions with sample sizes \(n_1 \in \{15, 16, \dots, 90\}\). For each sample size \(n_1\), the possible break positions are $T_{1,n_1} = \{10, \dots, n_1 - 5 \}$. Let \(\hat{T}_{n_1}^{(1)}\) be the estimated optimal single-break partition for sample size \(n_1\), and consider the collection \(\{\hat{T}_{n_1}^{(1)} \mid n_1 \in \{15, 16, \dots, 90\}\}\). In the second step, the candidate positions for the second break are \(T_2 \in \{15, 16, \dots, T - 10\}\), and each candidate \(T_2\) is associated with a corresponding optimal single-break partition \(\hat{T}^{(1)}_{T_2}\). The estimated two breaks are then obtained by selecting the pair \((\hat{T}^{(1)}_{T_2}, T_2)\) that produces the smallest sum of squared residuals among all two-break partitions.

It can be seen that the modified approach does not guarantee finding the global minimum, since the computation in the second step depends on the location of the single-break partition from the first step, which is not based on the full sample. In other words, the estimated location of the \(k\)-th break entirely depends on the locations of the previous \(k-1\) breaks, and not on the remaining \(m-k\) breaks. As a result, the algorithm is past-dependent and does not account for future observations. Nonetheless, this approach typically produces highly accurate estimates, although it occasionally fails to find the global minimizer. In such cases, the solution remains close to the optimal one, as will be shown in the simulations. In the next section, we discuss a method for determining the optimal number of breaks. This requires comparing models with different numbers of breaks, say, up to \(m\), which in turn requires the estimated partitions for models with \(k = 0,1, \dots, m\) breaks. This adds only a marginal increase in computational complexity relative to estimating the single optimal \(m\)-break partition, since information from the smaller subsamples can be reused when estimating the partitions for fewer breaks.

A modified DP algorithm for finding the optimal partitions with up to \(m\) structural breaks is defined as follows. Let \(SSR_n(\hat{T}^{(r-1)}_{T_r}, T_r)\) denote the sum of squared residuals obtained by OLS for a partition with \(r\) breaks in the first \(n\) observations, where \(\hat{T}^{(r-1)}_{T_r}\) represents the first \(r-1\) breaks corresponding to the optimal partition of the initial \(T_r\) observations, and the \(r\)-th break occurs at \(T_r\). For notational convenience, \((\hat{T}^{(0)}_{T_1}, T_1)\) denotes a single break at \(T_1\). The modified DP algorithm solves the optimization problem in three main steps:
\begin{enumerate}
    \item Initialize \(k = 1\). If \(k = m\), proceed directly to Step 2; otherwise, compute the subproblem solutions with \(k\) breaks:
    \[
    \hat{T}^{(k)}_{n_k} 
    = 
    \Bigl(
        \hat{T}^{(k-1)}_{T_k},\ 
        \arg\min_{T_k \in \{\, l_1 + (k-1)h_1, \dots, n_k - h_1 \,\}} 
        SSR_{n_k}\bigl(\hat{T}^{(k-1)}_{T_k}, T_k\bigr)
    \Bigr),
    \]
    where \(n_k\) is an admissible subsample size such that \(n_k \in \{\, l_1 + k h_1, \; l_1 + k h_1 + 1, \dots, T - l_2 \,\}\). The information from this step needs to be stored, as it will be used as input for the next iteration.

    \item Determine the optimal \(k\)-partition on the full sample by
    \begin{equation}
    \hat{T}^{(k)}_{T} 
    = 
    \Bigl(
        \hat{T}^{(k-1)}_{T_k},\ 
        \arg\min_{T_k \in \{\, l_1 + (k-1)h_1, \dots, T - l_2 \,\}} 
        SSR_T\bigl(\hat{T}^{(k-1)}_{T_k}, T_k\bigr)
    \Bigr).
    \label{eq:constrained dp}
    \end{equation}
    
    \item If \(k < m\), increase \(k\) by one and return to Step 1, using the information from Step 1 of the current iteration as input for Steps 1 and 2 of the next iteration; otherwise, if \(k = m\), stop. The solutions from Step 2 across all iterations then form the collection of optimal partitions with up to \(m\) breaks, \(\{\hat{T}_T^{(1)}, \dots, \hat{T}_T^{(m)}\}\).

\end{enumerate}

To obtain the solution for the modified DP algorithm, the seasonality regressors must be included in the computation of the objective function. However, information about the presence of seasonality is typically unavailable a priori. For simplicity, we therefore implement the algorithm without seasonality. Intuitively, this omission is not critical, as structural changes occur only in the trend component. The seasonality component merely captures oscillations around the trend and thus should have little effect on the estimated break locations. We verify this empirically by comparing the estimates obtained with and without the inclusion of seasonal regressors. 

\begin{figure}[t]
    \centering
    \includegraphics[width=0.9\linewidth]{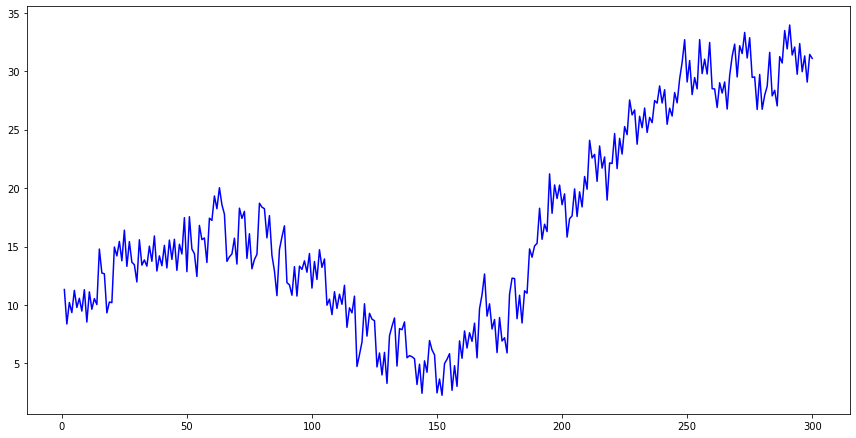}
    \caption{Sample realization of the specified data generating process for $T=300$.}
    \label{fig:dgp_300}
\end{figure}

\begin{figure}[t]
    \centering
    
    \begin{subfigure}[b]{0.3\textwidth}
        \includegraphics[width=\textwidth]{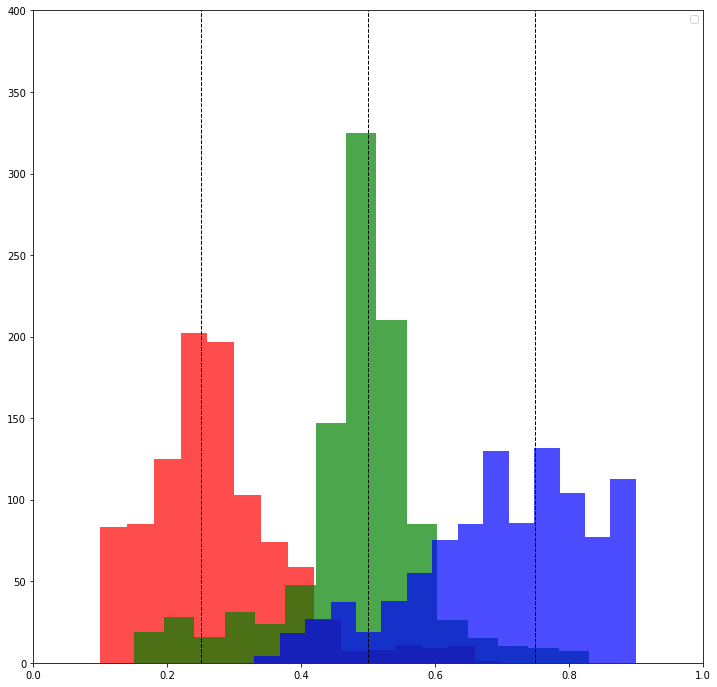}
    \end{subfigure}
    \begin{subfigure}[b]{0.3\textwidth}
        \includegraphics[width=\textwidth]{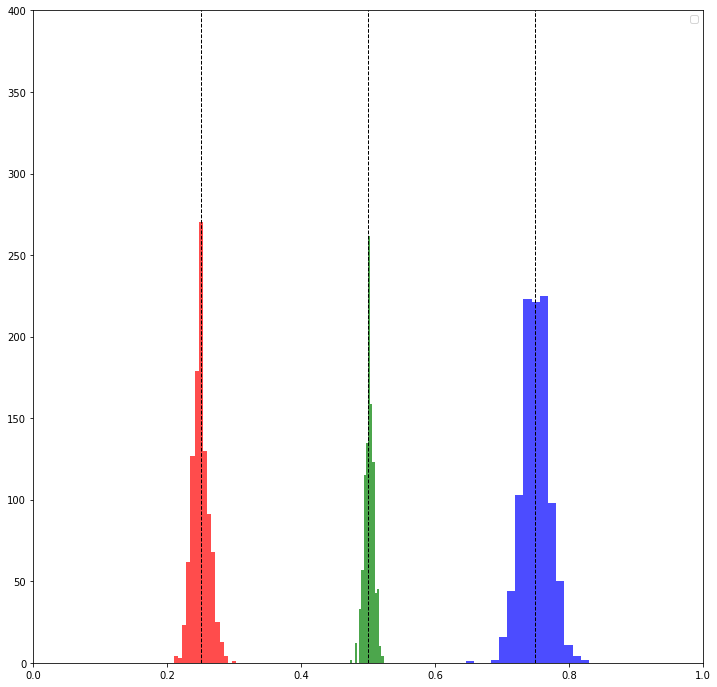}
    \end{subfigure}
    \begin{subfigure}[b]{0.3\textwidth}
        \includegraphics[width=\textwidth]{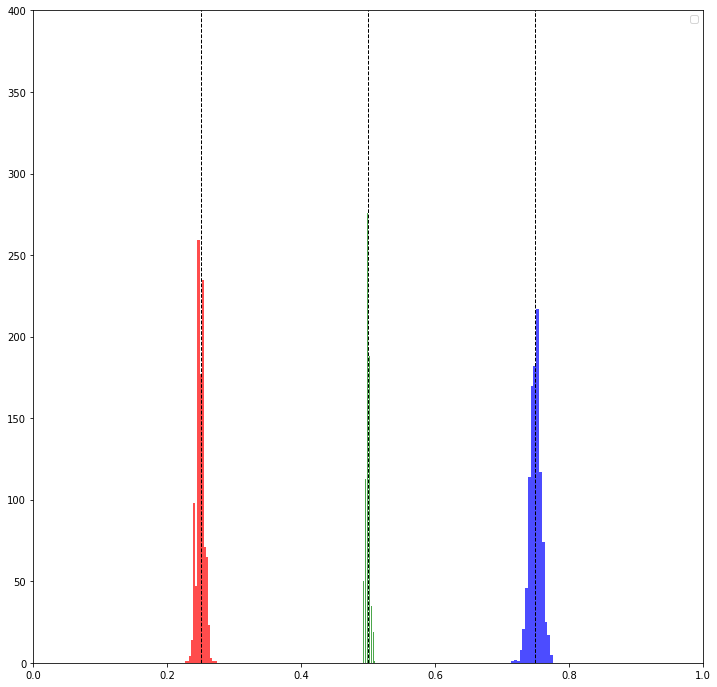}
    \end{subfigure}
    
    \begin{subfigure}[b]{0.3\textwidth}
        \includegraphics[width=\textwidth]{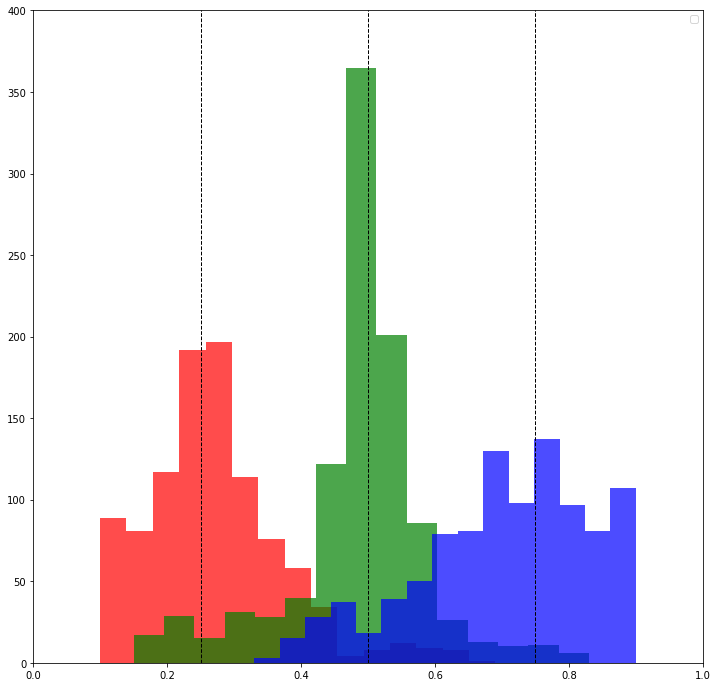}
    \end{subfigure}
    \begin{subfigure}[b]{0.3\textwidth}
        \includegraphics[width=\textwidth]{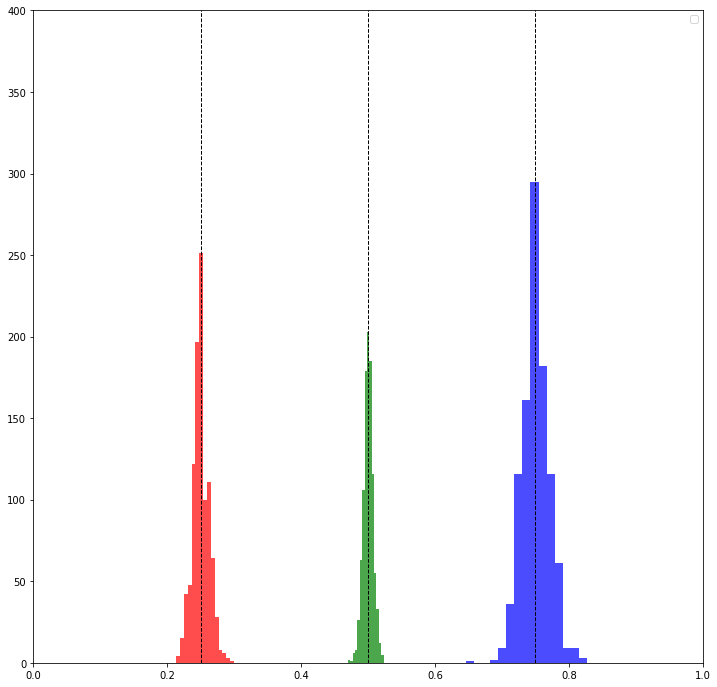}
    \end{subfigure}
    \begin{subfigure}[b]{0.3\textwidth}
        \includegraphics[width=\textwidth]{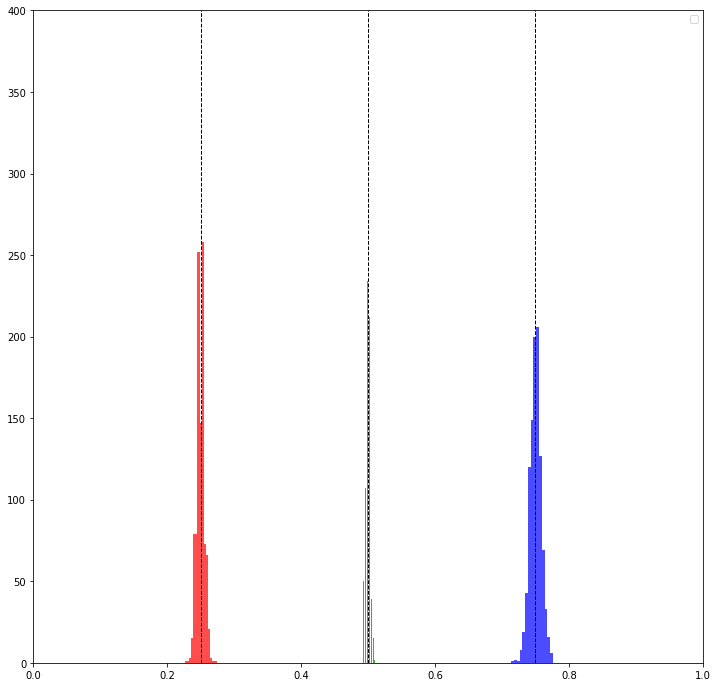}
    \end{subfigure}

    \begin{subfigure}[b]{0.3\textwidth}
        \includegraphics[width=\textwidth]{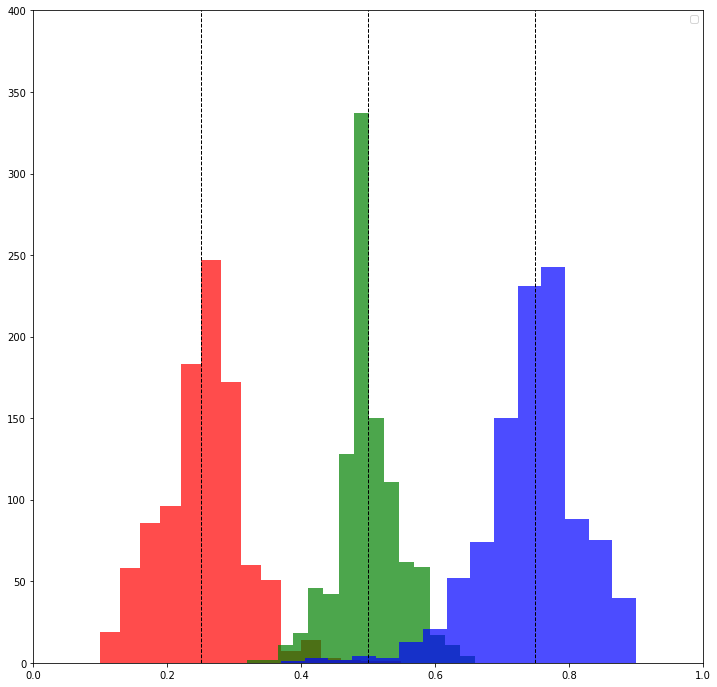}
    \end{subfigure}
    \begin{subfigure}[b]{0.3\textwidth}
        \includegraphics[width=\textwidth]{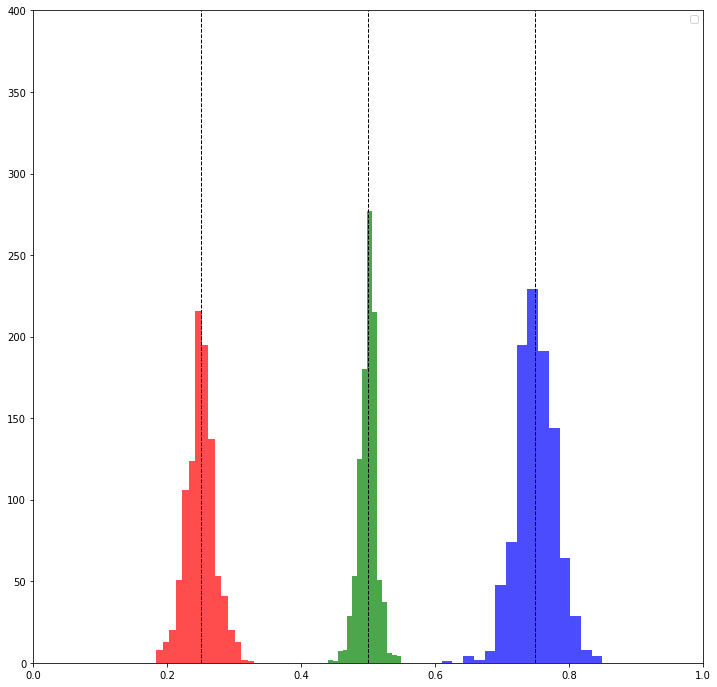}
    \end{subfigure}
    \begin{subfigure}[b]{0.3\textwidth}
        \includegraphics[width=\textwidth]{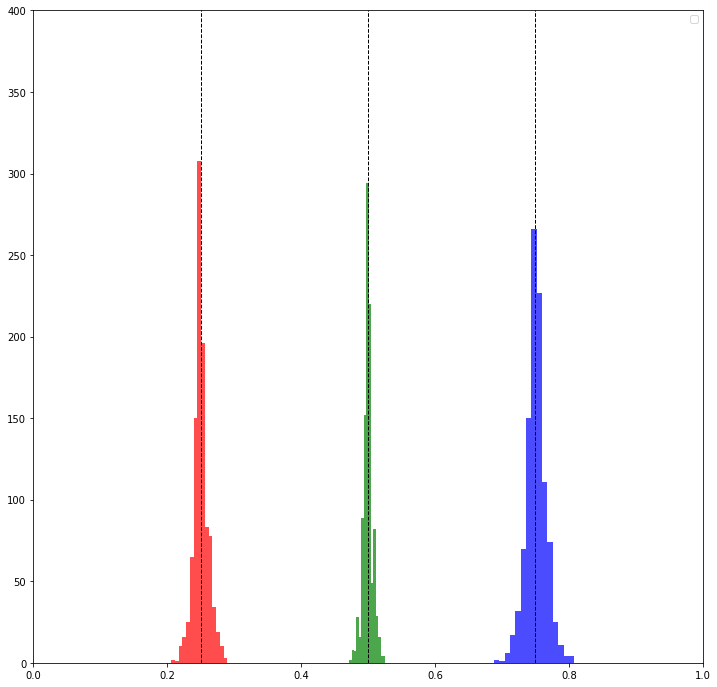}
    \end{subfigure}
    
    \caption{Histograms of the break fraction estimators. The columns, from left to right, correspond to sample sizes \(T=100\), \(300\), and \(500\). The top row panels show estimators with \(A_{T^{(m)}} = X_{T^{(m)}}\), the middle panels use \(A_{T^{(m)}} = [X_{T^{(m)}} \; Z]\), and the bottom row panels display the break fractions closest to the true values, selected from the estimates for \(m=6\) with \(A_{T^{(m)}} = X_{T^{(m)}}\). The dotted vertical lines indicate the true break fractions at \(\lambda_1 = 0.25\), \(\lambda_2 = 0.5\), and \(\lambda_3 = 0.75\).}
    
    \label{fig:break_fraction_estimator}
\end{figure}

From this point onward, the simulations are carried out using a data generating process (DGP) with the following specifications: the true break fractions are \(\lambda_1 = 0.25\), \(\lambda_2 = 0.5\), and \(\lambda_3 = 0.75\); the trend parameters are \(\mu_1 = 10\), \(\beta_1 = 0.1\), \(\beta_2 = -0.2\), \(\beta_3 = 0.3\), and \(\beta_4 = 0.1\); the seasonality parameters are \(\delta_1 = 1\), \(\delta_2 = -1.5\), \(\delta_3 = 0.75\), and \(\delta_4 = -0.25\); the error term follows an ARMA(1,1) process with \(\phi = 0.5\) and \(\theta = 0.5\), and the innovations are drawn from a normal distribution with variance \(\sigma_\epsilon^2 = 1\). Simulations are conducted for $T = 100$, $T = 300$ (refer to Figure~\ref{fig:dgp_300} for a sample realization), and $T = 500$, with 1000 replications for each sample size. The search for the breaks is restricted using the parameters \( h_1 = \lfloor 0.05T \rfloor \) and \( l_1 = l_2 = \lfloor 0.1T \rfloor \). Figure~\ref{fig:break_fraction_estimator} presents histograms of the break fraction estimators. 

Visual inspection (top and middle row panels) indicates that including the seasonality regressors in the computation of the global optimal solution does not noticeably alter the estimation accuracy. Furthermore, as the sample size increases, the estimated break fractions provided by the modified DP algorithm converge more closely to the true values, mimicking the behavior of a consistent estimator. It can also be seen from the bottom row panels that the estimates closest to the true values, selected from $\hat T^{(6)}_T$, remain highly accurate. This suggests that even when the model is overspecified, a subset of the estimated breaks can still be informative for inference. Intuitively, this is expected: when the model includes too many breaks such that \( m > m^0 \), where \( m^0 \) is the true number of breaks, the estimated break points tend to be placed so as to minimize the sum of squared residuals, causing them to cluster around the true break locations. As a result, the form of the estimated trend component remains approximately the same as the true trend component. This is a useful property, as it indicates that overspecification is not particularly detrimental and can still yield valuable information. More specifically, a subset of the estimated trend parameters, as well as the seasonality and ARMA parameters, should remain adequate even if the number of breaks is chosen to be larger than the true number of breaks.

In this paper, we assume that variation in the estimated break locations does not propagate to the estimation of the other model parameters. This assumption is justified since, as noted earlier, the break fractions converge at least at rate $T$, which is sufficiently fast to ensure that their estimation has little effect on the limiting distribution of the remaining model parameters, whose estimators converge at the standard $\sqrt{T}$ rate. Therefore, in the following procedures, we treat the estimated breaks of all partitions as known in order to determine the optimal number of breaks and obtain efficient estimates of the other model parameters. In other words, we view this problem as analogous to determining the appropriate set of regressors in multiple linear regression. For simplicity, confidence intervals for the estimated breaks are beyond the scope of this study; however, they may be important in certain applications and can be constructed using the methods described in \citet{bai_perron_2003} and \citet{ElliottMuller2007}.

\subsection{Estimating the Number of Breaks}

If the number of breaks \(m\) is known, the modified DP algorithm can be simplified by omitting the computation of optimal partitions for models with fewer breaks. More specifically, the range of sample sizes \(n_k\) in Step 1 can be slightly reduced at the end, and Step 2 is carried out only for the \(m\)-th iteration. However, in practice the number of breaks in a time series is unknown. Estimating this parameter is a nontrivial task, as the objective function always improves with an increasing number of breaks, necessitating alternative methods to determine the optimal specification. \citet{yao1988} recommends using the traditional Bayesian information criterion (BIC) to select the number of breaks, while \citet{liu1997} proposes a modified BIC, known as the LWZ information criterion, by imposing a much larger penalty. Under general conditions, both criteria provide consistent estimates of the number of breaks. \citet{Perron1997} conducted simulations using BIC and LWZ methods to select the number of breaks in a trend-shift model with serially correlated errors. He demonstrated that in the absence of autocorrelation in the errors, both BIC and LWZ provide accurate estimates, even when no breaks are present. However, when the errors are autocorrelated, BIC tends to systematically overestimate the number of breaks, whereas LWZ generally provides reliable estimates for moderate autocorrelation but struggles with highly autocorrelated errors. He also noted that the Akaike information criterion (AIC) provides extremely poor estimates, heavily overestimating the number of breaks, and should therefore not be used. 

\citet{Bai1997} and \citet{Chong2001} recommend a sequential testing procedure for determining the optimal number of breaks. The procedure begins by estimating a model with a small number of breaks deemed necessary and then tests whether including an additional break significantly improves the objective function. If the additional break is not significant, the procedure terminates and the current specification is selected as the optimal model; otherwise, the model includes one more break and the procedure is repeated until no further significant improvement is detected. \citet{bai_perron_1998} suggest that, unlike information-criteria-based methods, the sequential testing procedure provides an interpretable framework for comparing different models and can explicitly account for autocorrelation and heterogeneity in the error process. The consistent estimator can be achieved by appropriately adjusting the significance level of the test as the sample size increases. However, this approach is valid only for mean-shift or stationary regression models, as the break estimators remain consistent even when the model is underspecified \citep{Chong1995}.

\begin{figure}[t]
    \centering

    \begin{subfigure}[b]{0.3\textwidth}
        \includegraphics[width=\textwidth]{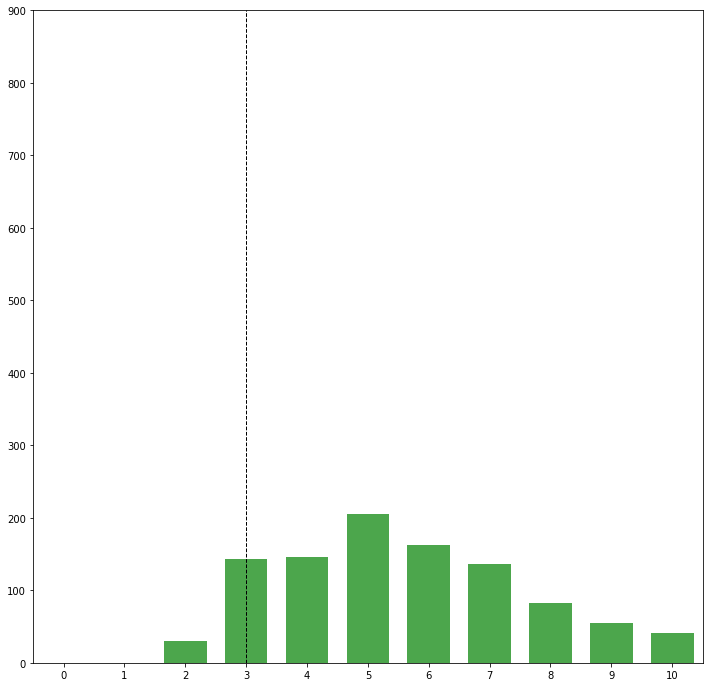}
    \end{subfigure}
    \begin{subfigure}[b]{0.3\textwidth}
        \includegraphics[width=\textwidth]{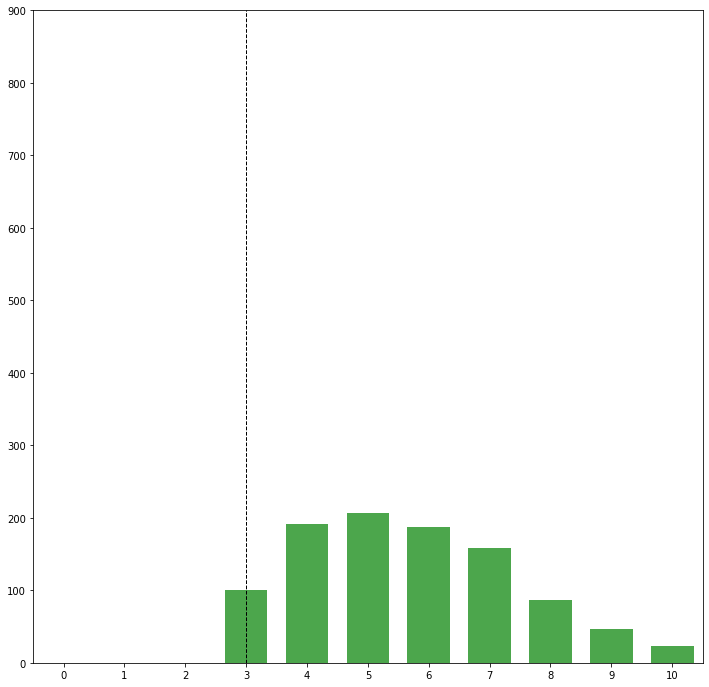}
    \end{subfigure}
    \begin{subfigure}[b]{0.3\textwidth}
        \includegraphics[width=\textwidth]{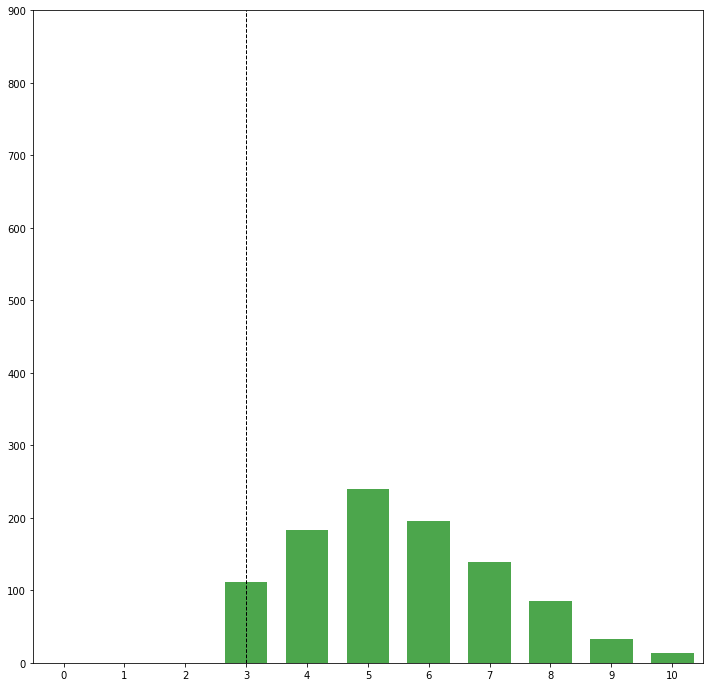}
    \end{subfigure}
    
    \begin{subfigure}[b]{0.3\textwidth}
        \includegraphics[width=\textwidth]{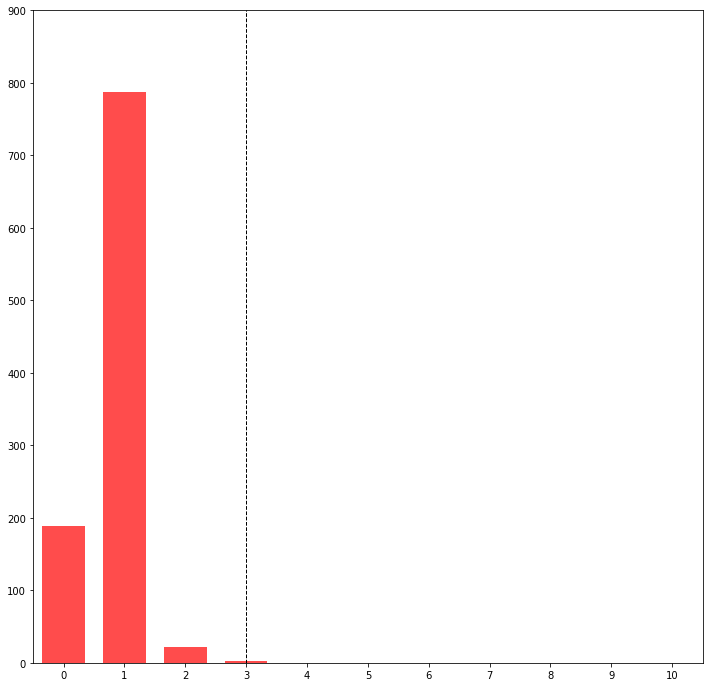}
    \end{subfigure}
    \begin{subfigure}[b]{0.3\textwidth}
        \includegraphics[width=\textwidth]{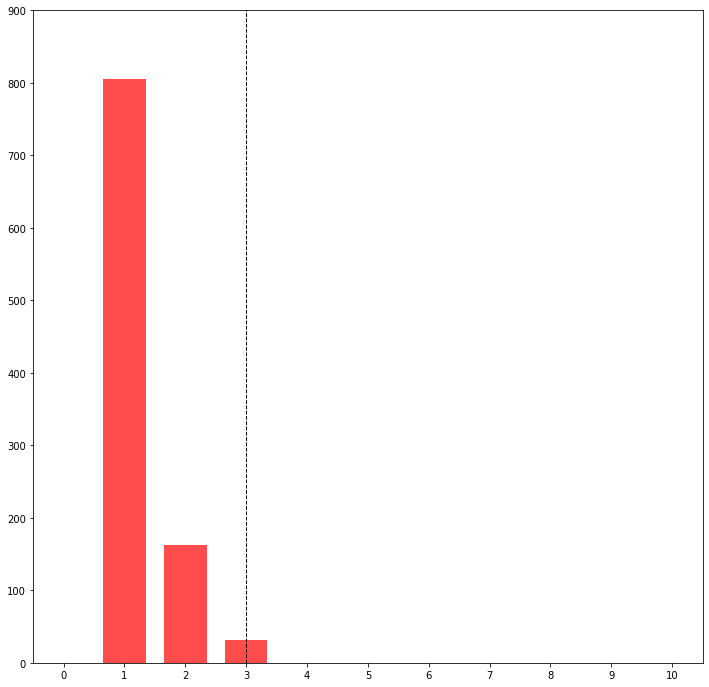}
    \end{subfigure}
    \begin{subfigure}[b]{0.3\textwidth}
        \includegraphics[width=\textwidth]{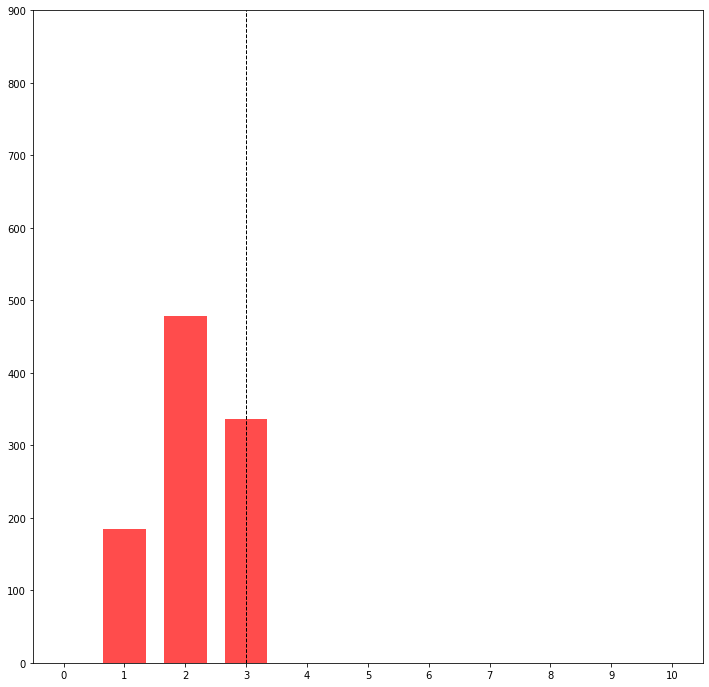}
    \end{subfigure}
    
    \caption{Histograms of the estimated number of breaks based on BIC (top) from the candidate models with up to 10 breaks, and ADF/KPSS tests (bottom) applied to the OLS residuals. From left to right: $T = 100$, $300$, and $500$. The dotted vertical lines indicate the true number of breaks.}

    \label{fig:num_bp_estimator1}
\end{figure}

In contrast, for trend-shift models, \citet{Yang2017} showed that underspecifying the number of breaks leads to inconsistent estimates, rendering the testing procedure invalid. This issue can be addressed by transforming the trend-shift model into an equivalent mean-shift model through differencing. However, this approach poses significant complications, since differencing stationary ARMA errors results in a non-invertible representation of the process. Therefore, it is only suitable when the error process is assumed to contain a unit root. To address the underspecification problem, we propose an alternative approach: instead of starting with a model with few breaks and adding more, we begin with a sufficiently large number $m_u$ and iteratively remove those models that are not statistically significant. As discussed earlier, the parameter estimates in an overspecified model remain meaningful, and thus this strategy effectively accounts for the issues that arise when models are underspecified. The choice of $m_u$ is determined according to BIC among the models with up to the specified $\bar m$ breaks. BIC is preferred in this case, as it tends to select an overspecified model and is a more established approach than LWZ, which applies a larger penalty and may underestimate the number of breaks in small samples. It is recommended to choose a sufficiently large $\bar m$ so that $\bar m > m^0$, ensuring that $m_u$ can be adequately determined; overall, setting $\bar m = 10$ works well for most time series. The top panels of Figure~\ref{fig:num_bp_estimator1} show that BIC generally selects an overspecified model for the same DGP specified in the previous section.

The central assumption of this paper is that nonstationarity can be represented as stationary fluctuations around deterministic trend and seasonal components once a sufficient number of structural breaks are introduced. At first glance, it may seem desirable to estimate the number of breaks by gradually increasing the number of imposed breaks, thereby reducing the persistence in the errors, until the residuals begin to mimic stationary behavior at some number of breaks, denoted by $m_l$. This strategy, however, tends to heavily underestimate the true number of breaks by interpreting the remaining deterministic patterns in the residuals as mean-reverting behavior similar to stationary dynamics. Nonetheless, it remains useful for establishing a lower bound on the number of breaks, since the methods discussed below assume that the errors are weakly stationary. Although not a formal estimation procedure, an approximate value of $m_l$ can be obtained by applying unit root or stationarity tests, such as the Augmented Dickey--Fuller (ADF; \citealp{SaidDickey1984}) and Kwiatkowski--Phillips--Schmidt--Shin (KPSS; \citealp{KPSS1992}) tests, to the OLS residuals corresponding to each candidate number of breaks, up to a previously determined $m_u$. Since the two tests have opposite null hypotheses, jointly considering them reduces ambiguity in borderline cases. The smallest number of breaks yielding strong evidence of stationary residuals can then be taken as $m_l$. The bottom panels of Figure~\ref{fig:num_bp_estimator1} show that the simultaneous use of ADF and KPSS tests generally struggles to determine the true number of breaks, although the accuracy slowly improves with sample size. The lag lengths for the tests are selected using BIC and the data-dependent method proposed by \citet{hobijn2004}, with critical values corresponding to significance levels of 0.01 and 0.1, respectively.

Consider a sample of observations $\{y_t\}_{t=1}^T$, assuming the absence of seasonality, with $m_u$ pre-estimated break locations $\hat{T}_T^{(m_u)} = (\hat{T}_1, \dots, \hat{T}_{m_u})$. We want to evaluate whether each pre-estimated break is statistically meaningful for these series, indicating a change in the underlying trend component. The analysis is based on the sequential prediction-interval-based algorithm, which provides a clear narrative behind the chosen model and is similar in nature to the cross-validation technique. More specifically, the procedure begins with the first break. Under the null hypothesis of no break or $\beta_1 = \beta_2$, the first two subsamples $Y_1 = (y_1, \dots, y_{\hat{T}_1})'$ and $Y_2 = (y_{\hat{T}_1+1}, \dots, y_{\hat{T}_2})'$ can be expressed as 
\begin{equation}
\begin{bmatrix}
Y_1 \\
Y_2
\end{bmatrix}
=
\begin{bmatrix}
X_1 & 0_{\hat T_1 \times 2} \\
0_{(\hat T_2 - \hat T_1) \times 2} & X_2
\end{bmatrix}
\begin{bmatrix}
\beta_1 \\
\beta_2
\end{bmatrix}
+ 
\begin{bmatrix}
U_1 \\
U_2
\end{bmatrix}
\label{eq:test}
\end{equation}
Here, $X_1 \in \mathbb{R}^{\hat T_1 \times 2}$ and $X_2 \in \mathbb{R}^{(\hat T_2 - \hat T_1) \times 2}$ are the matrices of time trend regressors without breaks, and $\beta_1$ and $\beta_2$ are the corresponding parameter vectors. The error vectors are $U_1 = (u_1, \dots, u_{\hat T_1})'$ and $U_2 = (u_{\hat T_1+1}, \dots, u_{\hat T_2})'$, so that $U = (U_1, U_2)'$ follows an ARMA process, with the pre-estimated parameter vector $(\hat{\phi}, \hat{\theta}, \hat{\sigma}^2_{\epsilon})'$, which is used to construct the theoretical autocovariance matrix of the process, $\hat{V}$ (discussed in the following section). As the errors are autocorrelated, the Feasible Generalized Least Squares (FGLS) estimator of $\beta_1$ based on the first subsample of observations is given by
\begin{equation}
\hat \beta_1=(X_1' \hat V_1^{-1} X_1)^{-1} X_1' \hat V_1^{-1} Y_1=\beta_1+(X_1' \hat V_1^{-1} X_1)^{-1} X_1' \hat V_1^{-1}U_1,
\label{eq:gls}
\end{equation}
where $\hat V_1=\hat V_{1:\hat T_1, 1:\hat T_1}$. 

We construct the difference between the vector $Y_2$ and its predicted values $\hat{Y}_{2|1}$, obtained from the regression estimated on the first subsample of observations. Specifically, the best linear unbiased predictor of $Y_2$, based on the matrix $\hat{V}_1$, is given by 
\begin{equation}
\hat{Y}_{2|1} = X_2 \hat{\beta}_1 + \hat R' \,  \hat V_1^{-1} (Y_1 - X_1 \hat{\beta}_1),
\label{eq:gls_predictor}
\end{equation}
where $\hat R = [\, \hat r_j \,]_{j=\hat T_1+1}^{\hat T_2}$ such that $\hat r_j = \bigl(\widehat{\mathrm{Cov}}(u_j, u_1),\ \dots,\ \widehat{\mathrm{Cov}}(u_j, u_{\hat{T}_1})\bigr)'$ denotes the column vector of theoretical covariances between the future error $u_j$ and the first subsample of errors $U_1$, constructed using the pre-estimated ARMA coefficients (see \citealp{Goldberger1962} for details). After applying Equation~\eqref{eq:gls}, the resulting difference $d = Y_2 - \hat{Y}_{2|1}$ can be written as
\begin{equation}
d =  X_2 \beta_2 + U_2 - X_2 \hat{\beta}_1 - \hat R' \, \hat V_1^{-1} (Y_1 - X_1 \hat{\beta}_1)
= X_2 \beta_2 + U_2 - X_2 \beta_1 - GU_1,
\label{eq: test_stat}
\end{equation}
where $G = (X_2 - \hat R' \hat V_1^{-1} X_1) (X_1' \hat V_1^{-1} X_1)^{-1}X_1' \hat V_1^{-1} + \hat R' \hat V_1^{-1}$.
The expectation of $d \in \mathbb{R}^{\hat{T}_2 - \hat{T}_1}$ is defined as
\begin{equation}
\mathbb{E}(d) = X_2 \beta_2 - X_2 \beta_1, 
\label{eq:expect_d}
\end{equation}
and the estimated covariance matrix of $d$ can be written as 
\begin{equation}
\widehat{\mathrm{Cov}}(d)
= [-G \;\; I_{\hat T_2- \hat T_1}\,]\hat V[-G \;\; I_{\hat T_2- \hat T_1}\,]' 
\label{eq:cov_d}
\end{equation}
Under the null hypothesis $\beta_2 = \beta_1$, the expression \eqref{eq:expect_d} is the zero vector. Consequently, if $\hat{V}$ is consistently estimated, the Wald-type test statistic can be expressed as
\begin{equation}
W = d' \, \big(\widehat{\mathrm{Cov}}(d)\big)^{-1} \, d,
\label{eq:wald_stat}
\end{equation}
with critical values obtained from a $\chi^2$ distribution with $\hat{T}_2 - \hat{T}_1$ degrees of freedom. Therefore, the break test is based on evaluating whether $d$ significantly deviates from zero, that is, whether the next subsample of observations can be adequately explained by the first subsample. When the estimated length of the first subsample is short, the estimation in \eqref{eq:gls} may be unreliable, which can lead to falsely detecting a significant first break; therefore, the minimization in \eqref{eq:constrained dp} is restricted to partitions satisfying $T_1 \geq l_1$. 

Unfortunately, standard asymptotic tests cannot be applied directly since the break is identified only under the alternative hypothesis, making the use of standard critical values inappropriate and biasing the test toward rejecting the null hypothesis. Consequently, the true distribution of the test statistic is stochastically larger and nonstandard. Comprehensive discussions of formal tests for structural change in stationary and nonstationary regression models with a time trend are provided by \citet{Andrews1993} and \citet{ChuWhite1992}, respectively. Nevertheless, it will be demonstrated that using standard critical values can still yield reasonable results, especially when the sample size is sufficiently large. This is because a stronger tendency to reject the null hypothesis leads to overspecification, which, as noted earlier, is generally less problematic than underspecification, thereby justifying this procedure.

\begin{figure}[t]
    \centering
    \includegraphics[width=0.9\linewidth]{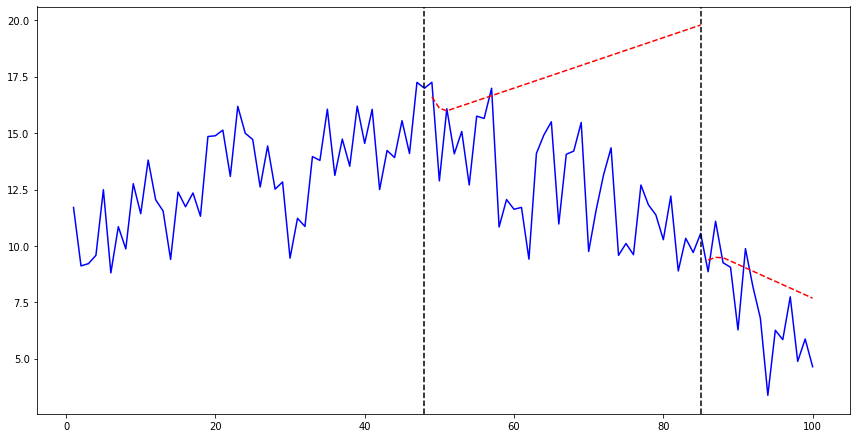}
    \caption{Artificially generated series with $T = 100$ and a true break at $T_1^0 = 50$. The dotted vertical lines indicate the locations of the two estimated breaks, and the dotted red line represents the sequential out-of-sample predictions.}
    \label{fig:break_example}
\end{figure}

If the null hypothesis of no break at $\hat{T}_1$ is rejected, we proceed to test the second pre-estimated break, $\hat{T}_2$, conditional on the presence of the first break at $\hat{T}_1$, which is incorporated into the trend regressors in \eqref{eq:test}. The procedure is identical to that described above. For example, consider an artificially generated series shown in Figure~\ref{fig:break_example}. We wish to evaluate whether the two-break model is adequate for this series. Visually, it can be seen that the first estimated break appears to be significant, as the first subsample cannot properly predict the second subsample. On the other hand, the third subsample can be adequately predicted by the first two segments, suggesting that the second estimated break may not be important. Therefore, this selection of the optimal model is very intuitive, as it simply evaluates whether the next subsample of observations originates from the same DGP as the previous subsamples. The testing framework we adopt in this setting helps determine appropriate thresholds that quantify the accuracy of the sequential out-of-sample predictions and assess whether they are acceptable.

If all estimated breaks within the $m_u$-partition are found to be significant, the model with $m_u$ breaks is considered optimal. Otherwise, if any of the breaks are found to be non-significant, the model is deemed suboptimal and a reduced model with $m_u-1$ pre-estimated breaks, denoted by $\hat{T}_T^{(m_u-1)}$, is considered by repeating the sequential testing of each break. This procedure is repeated until all pre-estimated breaks are found to be significant, in which case the model is considered optimally specified. For very large samples, such sequential analysis of different models is unnecessary, as spurious breaks have little effect on the estimated parameters and non-significant breaks from the $m_u$-partition can simply be discarded. However, for typical sample sizes, as discussed earlier, the estimated break locations are past-dependent, so a spurious break identified at the beginning may lead to inaccurate estimates of subsequent break locations. Therefore, we consider each model sequentially.

Consider a collection of pre-estimated breaks $\{\hat{T}_T^{(m_l)}, \dots, \hat{T}_T^{(m_u)}\}$. To determine the optimal number of breaks, $\hat{m}$, we use the following algorithm:
\begin{enumerate}
    \item Maximal model: Test the significance of each break in $\hat{T}_T^{(m_u)}$ sequentially.
    \begin{itemize}
        \item If all breaks are significant, set $\hat{m} = m_u$ and stop.
        \item If a non-significant break is detected, set $m = m_u - 1$ and proceed to Step~2.
    \end{itemize}

    \item Reduced models: Test the significance of each break in $\hat{T}_T^{(m)}$ sequentially.
    \begin{itemize}
        \item If all breaks are significant, set $\hat{m} = m$ and stop.
        \item If a non-significant break is detected and $m > m_l + 1$, decrease $m$ by one and repeat Step~2.
        \item If a non-significant break is detected and $m = m_l + 1$, set $\hat{m} = m_l$ and stop.
    \end{itemize}
\end{enumerate}
The algorithm terminates in two cases. First, if all specified breaks are significant, the procedure stops to avoid potential underspecification, even if subsequent models could also contain only significant breaks. Second, if the number of breaks reaches the minimum required for stationarity, the procedure stops, as it is no longer feasible when the errors are of unit root type. 

\begin{figure}[p]
    \centering
    
    \begin{subfigure}[b]{0.3\textwidth}
        \includegraphics[width=\textwidth]{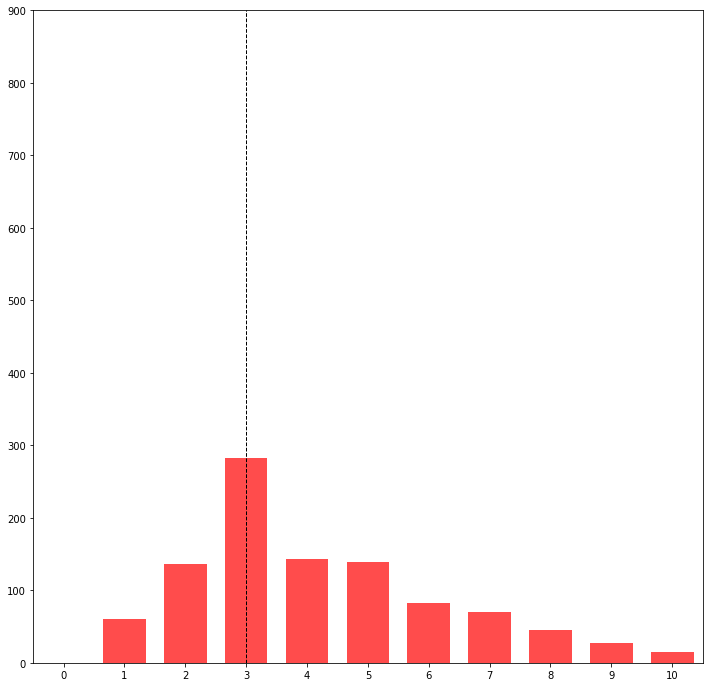}
    \end{subfigure}
    \begin{subfigure}[b]{0.3\textwidth}
        \includegraphics[width=\textwidth]{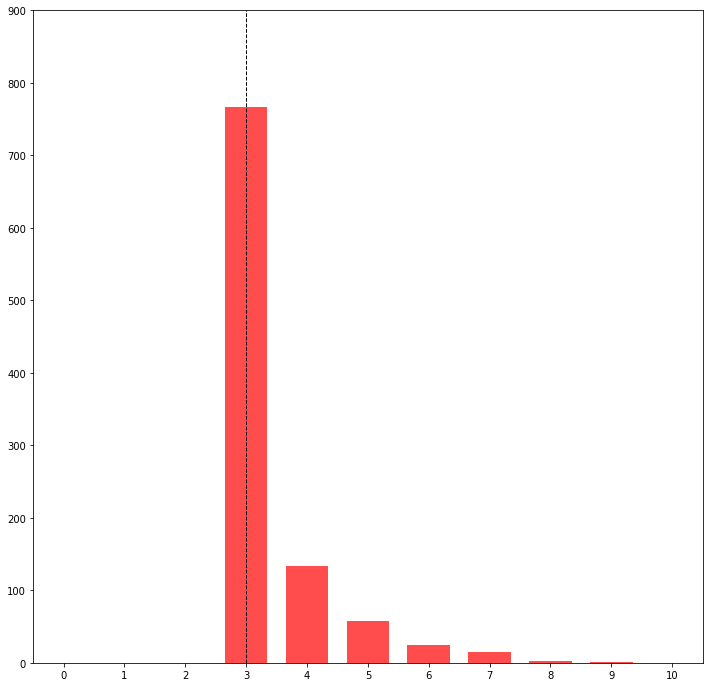}
    \end{subfigure}
    \begin{subfigure}[b]{0.3\textwidth}
        \includegraphics[width=\textwidth]{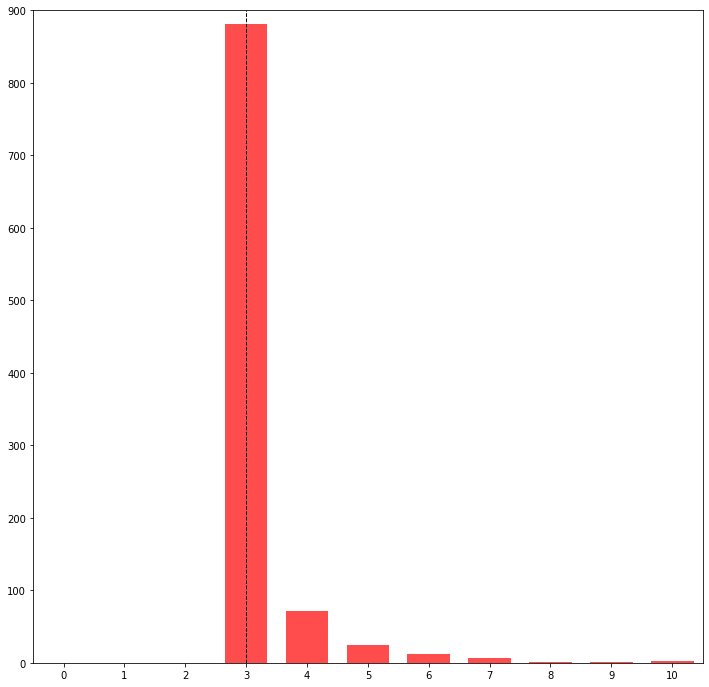}
    \end{subfigure}
    
    \begin{subfigure}[b]{0.3\textwidth}
        \includegraphics[width=\textwidth]{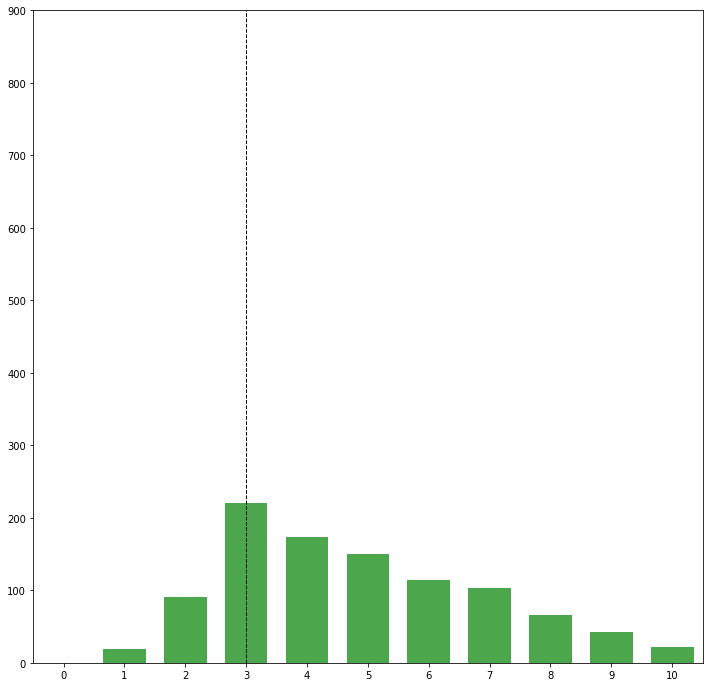}
    \end{subfigure}
    \begin{subfigure}[b]{0.3\textwidth}
        \includegraphics[width=\textwidth]{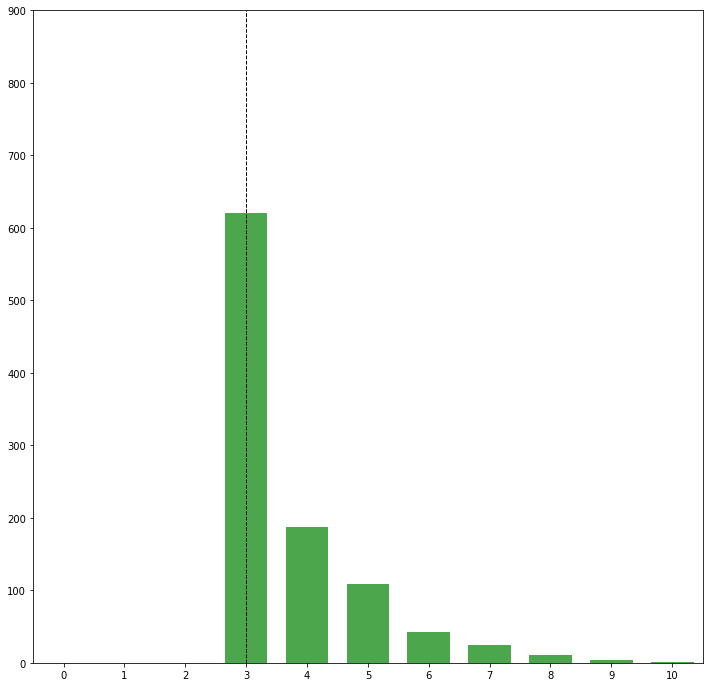}
    \end{subfigure}
    \begin{subfigure}[b]{0.3\textwidth}
        \includegraphics[width=\textwidth]{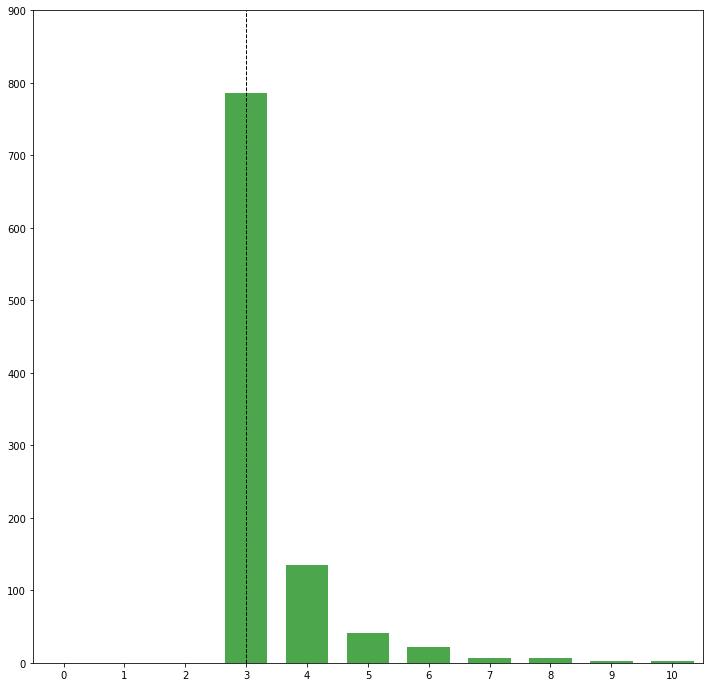}
    \end{subfigure}
    
    \begin{subfigure}[b]{0.3\textwidth}
        \includegraphics[width=\textwidth]{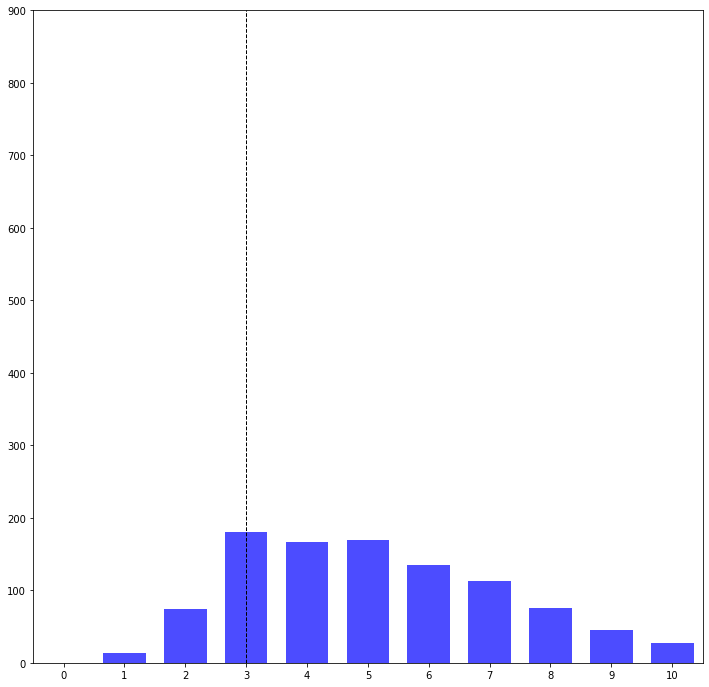}
    \end{subfigure}
    \begin{subfigure}[b]{0.3\textwidth}
        \includegraphics[width=\textwidth]{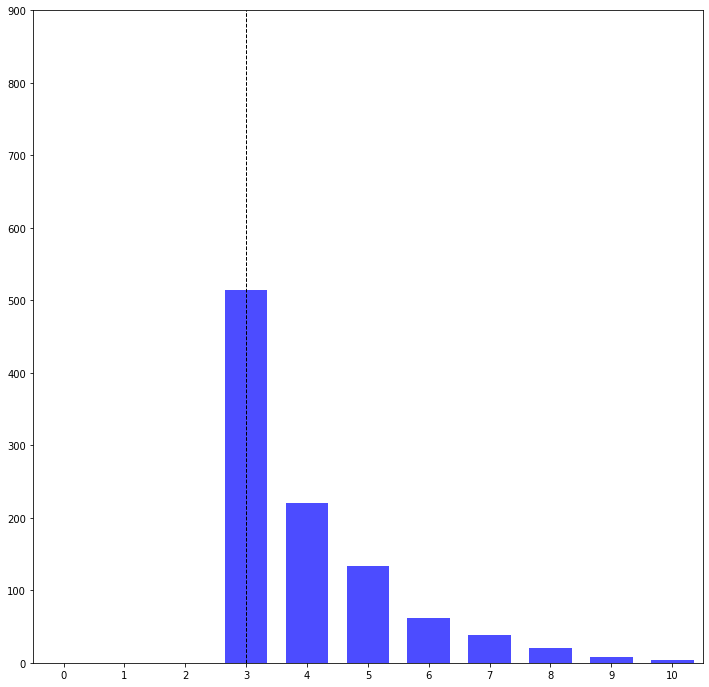}
    \end{subfigure}
    \begin{subfigure}[b]{0.3\textwidth}
        \includegraphics[width=\textwidth]{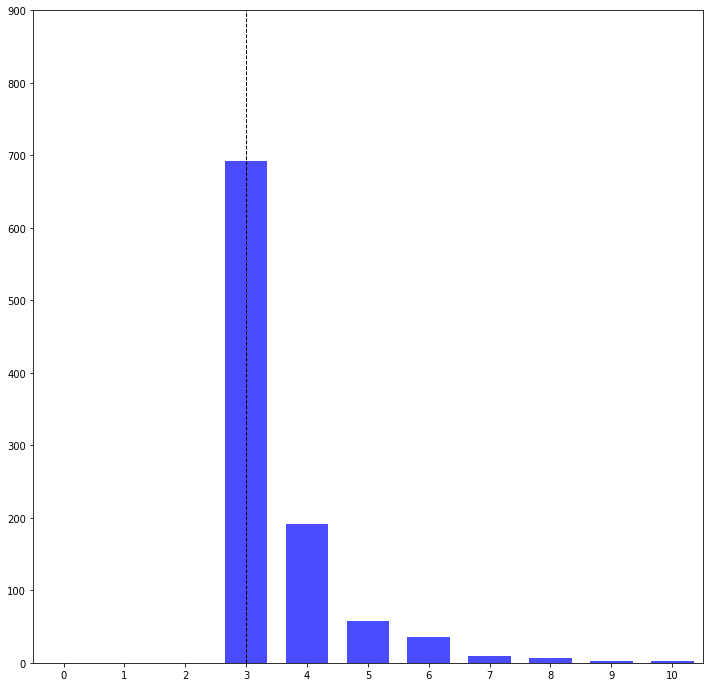}
    \end{subfigure}
    
    \begin{subfigure}[b]{0.3\textwidth}
        \includegraphics[width=\textwidth]{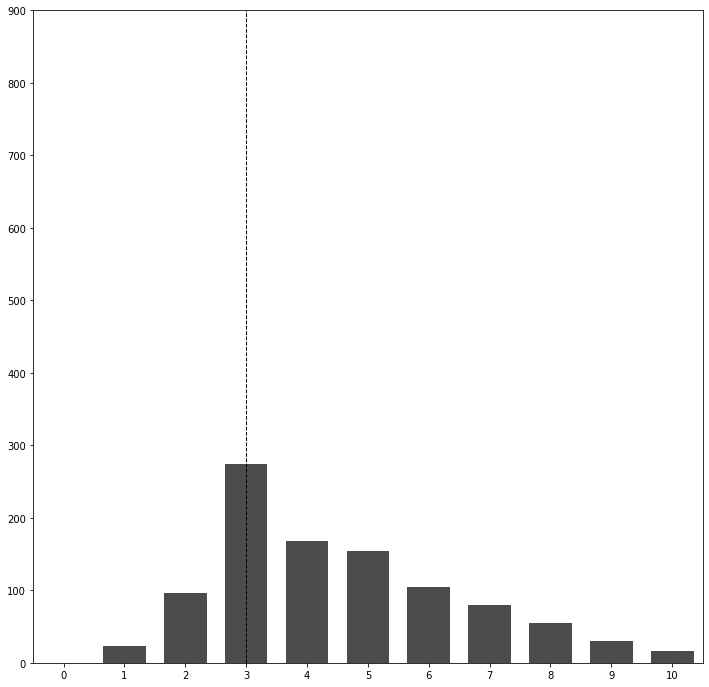}
    \end{subfigure}
    \begin{subfigure}[b]{0.3\textwidth}
        \includegraphics[width=\textwidth]{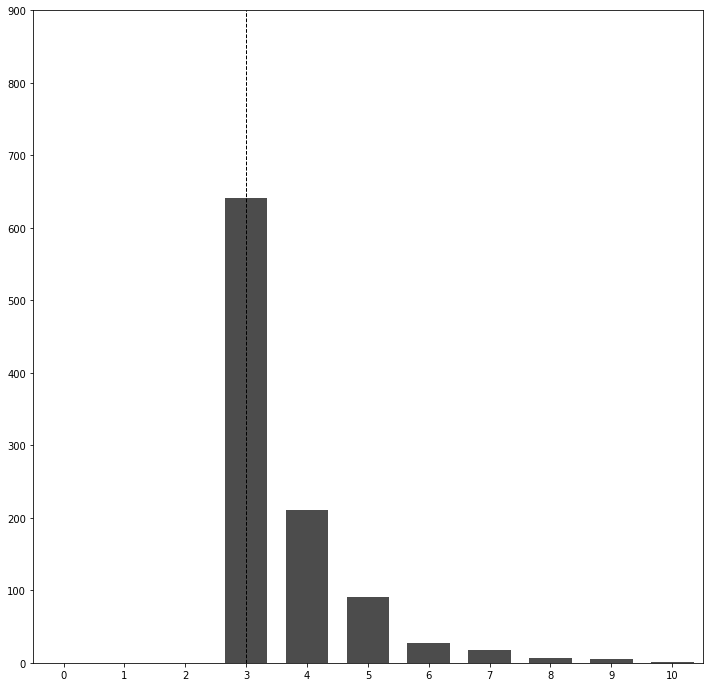}
    \end{subfigure}
    \begin{subfigure}[b]{0.3\textwidth}
        \includegraphics[width=\textwidth]{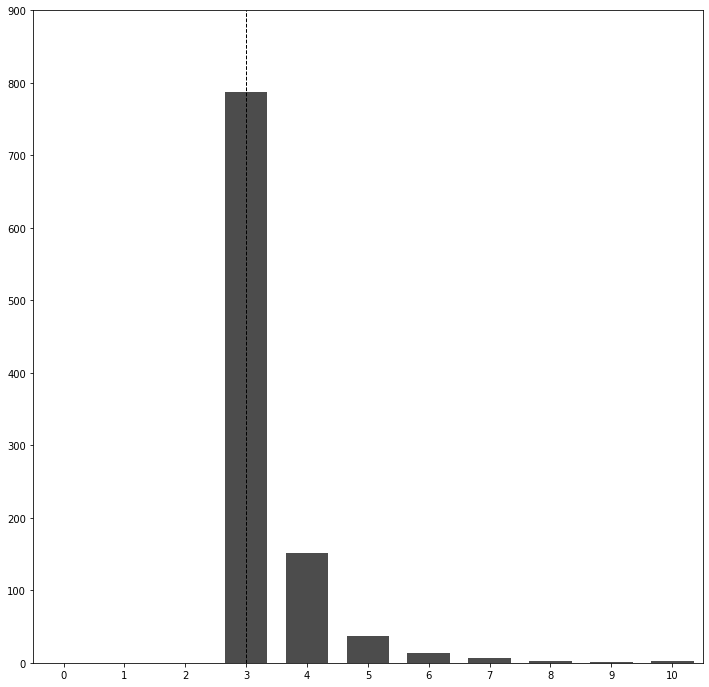}
    \end{subfigure}
    
    \caption{Histograms of the estimated number of breaks. From left to right: $T = 100$, $300$, and $500$. From top to bottom: estimators with typical significance levels (0.01, 0.05, and 0.1) and the adjusted estimator, respectively. The dotted vertical lines indicate the true number of breaks.}
    
    \label{fig:num_bp_estimator2}
\end{figure}

Importantly, the discussed algorithm, although based on a testing procedure, should not be regarded as a formal test for multiple breaks but rather as a model selection approach for determining the optimal number of breaks. The first three top rows of Figure~\ref{fig:num_bp_estimator2} present the simulation results for the proposed estimator of the number of breaks at typical significance levels of 0.01, 0.05, and 0.1, respectively, using the same series that were used for Figure~\ref{fig:num_bp_estimator1}. The construction of the matrices $\hat{V}$ and $\hat{R}$, based on the pre-estimated ARMA coefficients $(\hat{\phi}, \hat{\theta}, \hat{\sigma}^2_{\epsilon})'$ and the seasonality adjustment required for the sequential prediction-interval-based algorithm, is discussed in the next section. Simulation results demonstrate that, for moderate and large samples, the algorithm generally selects the true number of breaks across all specified significance levels $\alpha$. In contrast, for small samples, as $\alpha$ increases, underspecification occurs less often, but this simultaneously leads to more frequent overspecification.

In general, detection of breaks depends on both the magnitude of the change, $\beta_{i+1} - \beta_{i}$, and its duration, $\hat{T}_{i+1} - \hat{T}_{i}$. Since the estimation of the breaks relies on the squared loss function, the presence of unusually strong shocks may occasionally produce misleading results. Additionally, due to the autocorrelated nature of the errors, the system may be temporarily influenced, causing subsequent observations to fall outside the specified prediction interval and be incorrectly identified as significant breaks in short segments. In longer segments, such shocks tend to dissipate, and observations eventually return to within the prediction interval. To address this issue, a smaller significance level can be applied to short segments, increasing tolerance to abnormal short-term shocks. For convenience, a segment is considered short if $\hat{T}_i - \hat{T}_{i-1} \le h_2$ for $i = 1, \dots, m + 1$. The bottom row of Figure~\ref{fig:num_bp_estimator2} presents the simulation results using this adjustment with $h_2 = \lfloor 0.1T \rfloor$, showing that applying $\alpha_1 = 0.01$ to short segments to avoid overspecification and $\alpha_2 = 0.1$ to the remaining segments to guard against underspecification produces better results for short samples, while yielding highly accurate estimates for moderate and large samples. Overall, the sequential prediction-interval-based algorithm serves as an improvement over the BIC method when the errors are serially correlated. The proposed approach performs well in moderate and large samples, providing a number of breaks at least equal to the true value and, with appropriate adjustments, can still yield adequate estimates in small samples.

\subsection{Full Model Estimation}

The estimation of the number of breaks relies on the presence or absence of the seasonal component and the autocorrelation structure of the errors, which are in practice unknown. Conventional OLS estimates cannot be directly used to test the null hypothesis of no seasonality, $\delta = 0$, as their covariance matrix is inconsistent under potential serial correlation. A common remedy is to employ a heteroskedasticity and autocorrelation consistent (HAC) estimator, which provides a valid covariance matrix of the parameter estimates and allows standard inference procedures such as the Wald test. To assess autocorrelation in the errors, the Ljung--Box test can be applied to the OLS residuals. Both procedures are implemented in accordance with the methods proposed by \citet{newey1987simple} and \citet{ljung1978measure}. Both procedures can be used with an estimated $m_u$-partition, since the seasonal parameters and residuals remain meaningful for overspecified models. If the null hypothesis of no autocorrelation cannot be rejected, the estimated number of breaks is selected using BIC, which yields accurate estimates under uncorrelated errors.

For serially correlated errors, once the specification of the seasonal component is established, the seasonal coefficients must be estimated to seasonally adjust the observations—that is, by subtracting the fitted seasonal component—and the ARMA coefficients must be estimated to construct the matrices $\hat{V}$ and $\hat{R}$, which are subsequently used in the prediction-interval-based algorithm. We estimate these parameters conditional on the $m_u$-partition and use them in the sequential testing procedure of each model. The preliminary parameter estimates are obtained iteratively by combining OLS and FGLS estimation for the trend and seasonal parameters with the Hannan--Rissanen (HR) algorithm \citep{hannan1982recursive} for estimating the specified ARMA coefficients. The estimation proceeds as follows:
\begin{enumerate}
    \item Estimate the trend and seasonal parameters using OLS, and compute the residuals.
    \item Estimate the ARMA coefficients from the OLS (or FGLS) residuals using the Hannan--Rissanen (HR) method.
    \item Construct the theoretical covariance matrix of the error process from the ARMA estimates in Step~2, and re-estimate the trend and seasonal parameters via FGLS to obtain updated residuals.
    \item Repeat Steps~2 and~3 until the parameter estimates converge.
\end{enumerate}
The orders of the ARMA process are chosen based on BIC, due to its asymptotic consistency, which is required for the sequential prediction-interval-based algorithm. Accordingly, the algorithm is executed for all candidate combinations of $p$ and $q$ with $p \leq p_{\max}$ and $q \leq q_{\max}$, and the ARMA specification yielding the lowest BIC is selected. The corresponding preliminary estimates of $\hat{\delta}$ and $(\hat{\phi}, \hat{\theta}, \hat{\sigma}^2_{\epsilon})'$ are then used to determine the optimal number of breaks.

Importantly, the discussed testing framework for the breaks does not account for the variability in the seasonality adjustment or in the pre-estimated ARMA parameters. However, since the uncertainty primarily arises from the white noise term, any underestimation in \eqref{eq:cov_d} would only slightly bias the tests toward rejecting the null hypothesis, which, as discussed earlier, does not pose substantial problems. One could account for the variation in the preliminary parameter estimates by jointly estimating all parameters within subsamples. Nonetheless, since the seasonality and ARMA coefficients are assumed to be constant, we can use the full sample for efficiency and avoid iterations and potential issues associated with parameter estimation in small samples. The simulation results shown in Figure~\ref{fig:num_bp_estimator2} were obtained with $p_{\max} = q_{\max} = 3$. Finally, once the number of breaks is determined, the order of the ARMA process is re-estimated conditional on the $\hat{m}$-partition $\hat T^{(\hat m)}_T$, after which the trend, seasonality, and ARMA parameters are re-estimated using standard numerical procedures to obtain Quasi-MLE estimates, which are robust and asymptotically efficient. 

Although the seasonal–trend–stationary ARMA model (STSA) is primarily designed for inference—particularly for analyzing how the trend evolves in response to various events—it can also be employed for forecasting. Reliable forecasts, however, can be obtained only when no structural shifts are expected within the forecast horizon; hence, the model is best suited for short-term forecasting. Out-of-sample predictions from the STSA model are obtained by extrapolating the most recently estimated regime of the trend, seasonal, and ARMA components into the future. In some cases, the model may incorrectly detect a structural break near the end of the sample, rendering the estimates of the final regime unstable due to the small size of the segment. To mitigate this issue, the candidate location of the last break is restricted such that $T_m \leq T - l_2$, analogous to the constraint on the initial segment size. 

Overall, parameters for the break search and model selection can be set based on the series' background information and the specific objectives. When such information is not available, we recommend using the following values: $h_1 = \lfloor 0.05T \rfloor$, $l_1=l_2=h_2 = \lfloor 0.1T \rfloor$, $\bar{m} = 10$, $\alpha_1 = 0.01$, $\alpha_2 = 0.1$, $p_{\max} = 3$, and $q_{\max} = 3$. The proposed model can be viewed as an extension of the Autoregressive Moving Average with Exogenous Regressors (ARMAX) model, where the exogenous components are represented by the trend and seasonal terms. The estimation procedures are implemented using the \texttt{statsmodels} library in Python \citep{seabold2010statsmodels}. The following section discusses several applications of the STSA model for both inference and forecasting.

\section{Empirical Applications}

\subsection{U.S Exports of Goods to Mainland China}

\begin{figure}[t]
    \centering
    \includegraphics[width=0.9\linewidth]{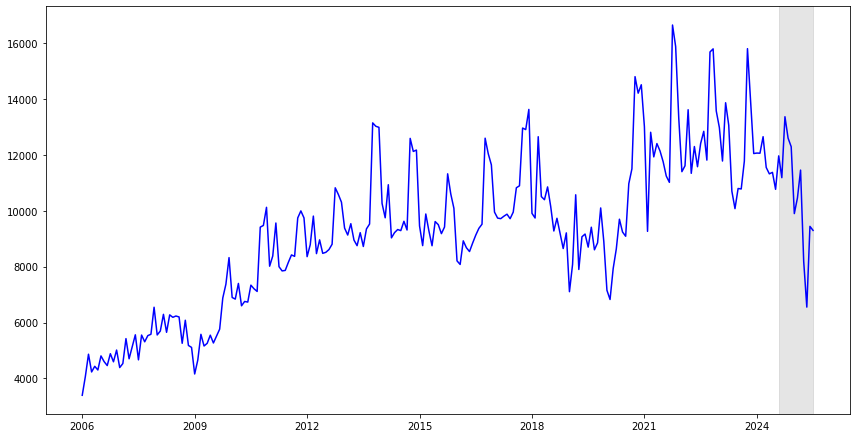}
    \caption{U.S. exports of goods to Mainland China series (January 2006 – July 2025). The 1-year interval shaded in gray indicates the test sample.}
    \label{fig:retail_sales}
\end{figure}

For our empirical illustration, we analyze the monthly U.S. exports of goods to Mainland China (in millions of dollars), covering the period from January 2006 to July 2025 \citep{us_census_export_2025}. Figure~\ref{fig:retail_sales} presents the series divided into a training sample (January 2004 – July 2024) for parameter estimation and a test sample (August 2024 – July 2025) for evaluating the forecasting performance of the STSA. To stabilize the variance, a logarithmic transformation is applied to the original series; the analysis is conducted on the transformed series without loss of generality. The series exhibits pronounced trend and seasonal patterns, making it suitable for evaluating the STSA framework’s ability to capture business-cycle dynamics driven by geopolitical tensions, trade disputes, global crises, and other disruptions. In this case, the parameters of the STSA models are specified according to the recommended values described earlier.

\begin{figure}[p]
    \centering
    \includegraphics[width=0.9\linewidth]{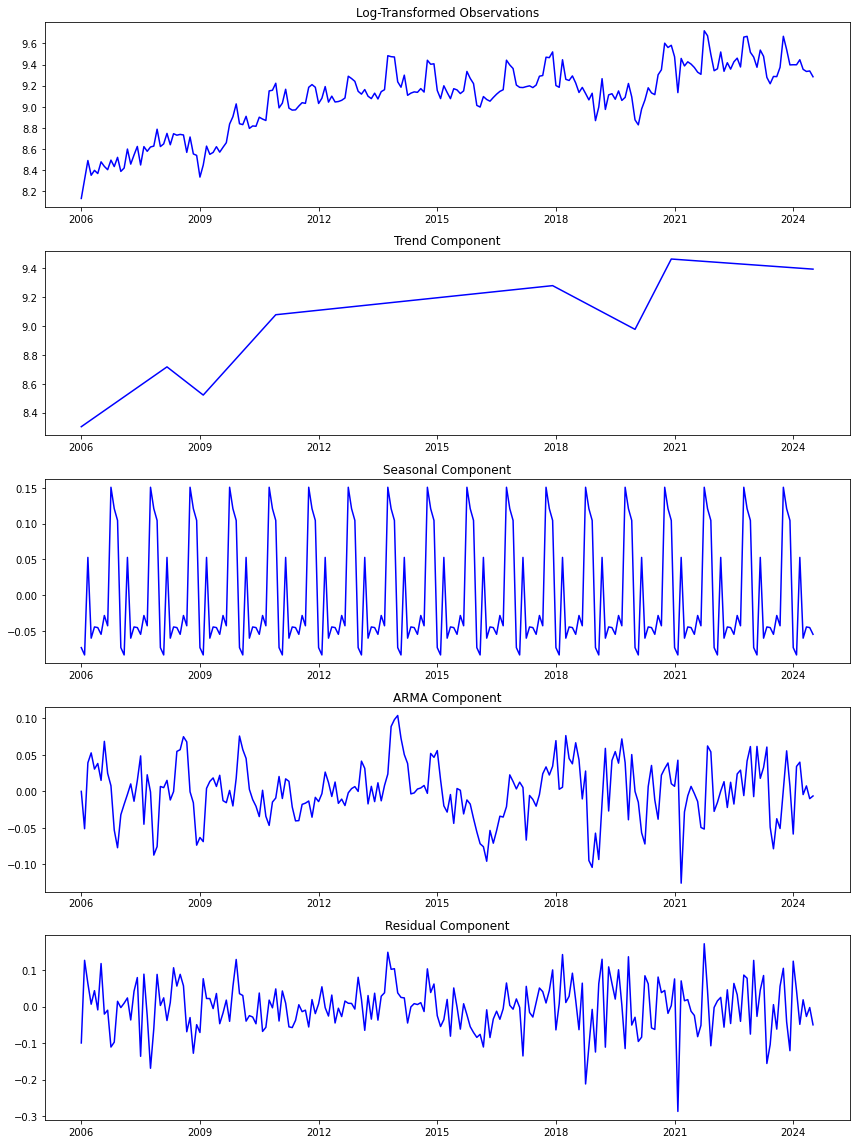}
    \caption{Decomposition of the observed series into the fitted trend, seasonal, and ARMA components, along with the residuals.}
    \label{fig:retail_decomposition}
\end{figure}

The intermediate results of the model specification and selection procedures are reported in Table~\ref{tab:intermediate_results}. For the upper and lower candidates for the number of breaks, BIC selects $m_u = 8$, while the ADF/KPSS tests yield $m_l = 0$, respectively. Tests for the absence of seasonality and autocorrelation in the residuals are strongly rejected for the $m_u$ partition. Candidate models with 8 and 7 breaks were systematically discarded, as the null hypothesis for the fourth break could not be rejected for the both cases. The sequential prediction-interval-based algorithm terminates at $m = 6$, identifying significant breaks in March~2008, February~2009, December~2010, December~2017, January~2020, and December~2020. Notably, the estimated break dates in the selected model also appear in the models with 7 and 8 breaks. Moreover, while BIC selects $m_u = 8$ breaks by identifying additional breaks around 2014 and 2016, our proposed algorithm indicates that these breaks are not sufficiently strong and therefore not relevant to include in the trend component, as can be visually seen from the top panel of Figure~\ref{fig:retail_decomposition}. 

\begin{table}[t]
\centering
\footnotesize
\begin{threeparttable}
\caption{Intermediate Results of STSA Model\protect\footnotemark[1]}
\label{tab:intermediate_results} 
\begin{tabular}{c c c c c c c c}
\hline
\noalign{\vskip 4pt}
$m_u$ & $m_l$ & WS\protect\footnotemark[2] & LB\protect\footnotemark[3] & & & & \\ 
8 & 0 & $211.89^{***}$ & $125.01^{***}$ & & & & \\[4pt]
\hline
\noalign{\vskip 4pt}
\multicolumn{8}{c}{$m=8$} \\
\noalign{\vskip 4pt}
$\hat T_1$ & $\hat T_2$ & $\hat T_3$ & $\hat T_4$ & $\hat T_5$ & $\hat T_6$ & $\hat T_7$ & $\hat T_8$ \\ 
03/2008 & 02/2009 & 11/2010 & 11/2014 & 03/2016 & 11/2017 & 01/2020 & 12/2020 \\
$37.92^{***}$ & $41.89^{***}$ & $63.78^{*}$ & $21.09$ (0.17) & & & & \\[4pt]
\hline
\noalign{\vskip 4pt}
\multicolumn{8}{c}{$m=7$} \\
\noalign{\vskip 4pt}
$\hat T_1$ & $\hat T_2$ & $\hat T_3$ & $\hat T_4$ & $\hat T_5$ & $\hat T_6$ & $\hat T_7$ & \\ 
03/2008 & 02/2009 & 12/2010 & 01/2017 & 12/2017 & 02/2020 & 01/2021 \\
$37.92^{***}$ & $46.31^{***}$ & $143.41^{***}$ & $10.16$ (0.52) & & & & \\[4pt]
\hline
\noalign{\vskip 4pt}
\multicolumn{8}{c}{$m=6$} \\
\noalign{\vskip 4pt}
$\hat T_1$ & $\hat T_2$ & $\hat T_3$ & $\hat T_4$ & $\hat T_5$ & $\hat T_6$ & & \\ 
03/2008 & 02/2009 & 12/2010 & 12/2017 & 01/2020 & 12/2020 & \\
$37.92^{***}$ & $46.31^{***}$ & $155.60^{***}$ & $88.91^{***}$ & $75.06^{***}$ & $146.48^{***}$ & & \\[4pt]
\hline
\end{tabular}
\begin{tablenotes}[flushleft]
\footnotesize
\item \textit{Notes:}
\item[1] $***$, $**$, and $*$ denote rejection of the null hypothesis at the 1\%, 5\%, and 10\% significance levels, respectively. Values in parentheses indicate $p$-values for which the null hypothesis cannot be rejected. The test statistic for the no-break hypothesis is reported under each tested break.
\item[2] The Wald test based on the HAC covariance matrix is employed to test the null hypothesis of no seasonality.
\item[3] The Ljung–Box test for absence of autocorrelation was conducted up to lag 10.
\end{tablenotes}
\end{threeparttable}
\end{table}

Table~\ref{tab:final_results} presents the summary results of the selected model, while Figure~\ref{fig:retail_decomposition} displays the decomposition of the U.S. exports of goods to Mainland China series into its estimated trend, seasonal, and ARMA components, along with the residuals. The estimated break dates correspond to major economic and geopolitical events that substantially affected the trade dynamics. Exports grow steadily before 2008, reflecting strong bilateral trade, after which the trend turns sharply negative during the global financial crisis (2008–2009). A strong rebound follows through 2010, supported by China’s large stimulus program and the global recovery, after which growth slows to nearly zero from 2010 to 2017. Beginning in late 2017, the trend declines again as U.S.–China trade tensions and tariff actions introduce uncertainty and suppress export growth. A brief but sharp upward shift appears in early 2020, corresponding to the Phase One trade deal, under which China committed to increasing purchases of U.S. goods, as well as temporary surges in demand during the initial COVID-19 disruptions. After late 2020, however, the trend flattens, reflecting persistent geopolitical frictions, the aftereffects of the pandemic, and the gradual diversification of global supply chains. Overall, the trend exhibits two V-shaped patterns, each followed by periods of new growth rates that reflect changes in economic conditions and trade dynamics.

The seasonal coefficients reveal a clear pattern in U.S. exports to Mainland China. Exports are below average from January to September, with especially weak activity early in the year due to post-holiday slowdowns and Lunar New Year–related disruptions on the Chinese side. Beginning in October, exports rise sharply and remain elevated through December, driven by stronger year-end demand, inventory restocking, and increased pre-holiday shipping. Overall, trade activity consistently peaks in the fourth quarter and bottoms out in the first quarter. For the error component, BIC selects an AR(1) stochastic process, indicating moderate serial dependence and suggesting that short-term shocks in trade activity tend to persist briefly before dissipating—a behavior commonly observed in aggregated macroeconomic time series.

\begin{table}[t]
\centering
\footnotesize
\begin{threeparttable}
\caption{Summary Results of STSA Model\protect\footnotemark[1]}
\label{tab:final_results} 
\begin{tabular}{c c c c c c c c}
\hline
\noalign{\vskip 4pt}
\multicolumn{8}{c}{Break Parameters} \\
\noalign{\vskip 4pt}
$\hat m$ & $\hat T_1$ & $\hat T_2$ & $\hat T_3$ & $\hat T_4$ & $\hat T_5$ & $\hat T_6$ & \\ 
6 & 03/2008 & 02/2009 & 12/2010 & 12/2017 & 01/2020 & 12/2020 & \\[4pt]
\hline
\noalign{\vskip 4pt}
\multicolumn{8}{c}{Trend Component Parameters} \\
\noalign{\vskip 4pt}
$\hat \mu_1$ & $\hat \beta_1$ & $\hat \beta_2$ & $\hat \beta_3$ & $\hat \beta_4$ & $\hat \beta_5$ & $\hat \beta_6$ & $\hat \beta_7$ \\ 
8.2870 & 0.0159 & -0.0178 & 0.0253 & 0.0024 & -0.0121 & 0.0443 & -0.0016 \\
(0.0562) & (0.0032) & (0.0058) & (0.0020) & (0.0004) & (0.0024) & (0.0069) & (0.0018) \\[4pt]
\hline
\noalign{\vskip 4pt}
\multicolumn{8}{c}{Seasonality Component Parameters\protect\footnotemark[2]} \\
\noalign{\vskip 4pt}
$\hat \delta_1$ & $\hat \delta_2$ & $\hat \delta_3$ & $\hat \delta_4$ & $\hat \delta_5$ & $\hat \delta_6$ & $\hat \delta_7$ & $\hat \delta_8$ \\ 
-0.0728 & -0.0830 & 0.0529 & -0.0596 & -0.0440 & -0.0449 & -0.0542 & -0.0280 \\
(0.0158) & (0.0181) & (0.0166) & (0.0128) & (0.0146) & (0.0132) & (0.0136) & (0.0142) \\
$\hat \delta_9$ & $\hat \delta_{10}$ & $\hat \delta_{11}$ & $\hat \delta_{12}$ & & & & \\
-0.0423 & 0.1506 & 0.1208 & 0.1044 & & & & \\
(0.0118) & (0.0217) & (0.0177) & & & & & \\[4pt]
\hline
\noalign{\vskip 4pt}
\multicolumn{8}{c}{Error Component Parameters} \\
\noalign{\vskip 4pt}
$p$ & $q$ & $\hat \phi$ & $\hat \sigma^2_{\epsilon}$ & & & & \\ 
1 & 0 & 0.5131 & 0.0046 & & & & \\
& & (0.0625) & (0.0005) & & & & \\[4pt]
\hline
\end{tabular}
\begin{tablenotes}[flushleft]
\footnotesize
\item \textit{Notes:}
\item[1] Quasi-maximum likelihood covariance matrix used for robustness to some misspecifications; calculated using the observed information matrix (complex-step) described in \citet{harvey1989forecasting}. Standard errors of the parameter estimates are reported in parentheses.
\item[2] The seasonal effect for December is computed under the constraint that the sum of all seasonal coefficients is zero: $\sum_{j=1}^{P} \hat{\delta}_j = 0$.
\end{tablenotes}
\end{threeparttable}
\end{table}

\begin{figure}[t]
    \centering
    \includegraphics[width=0.9\linewidth]{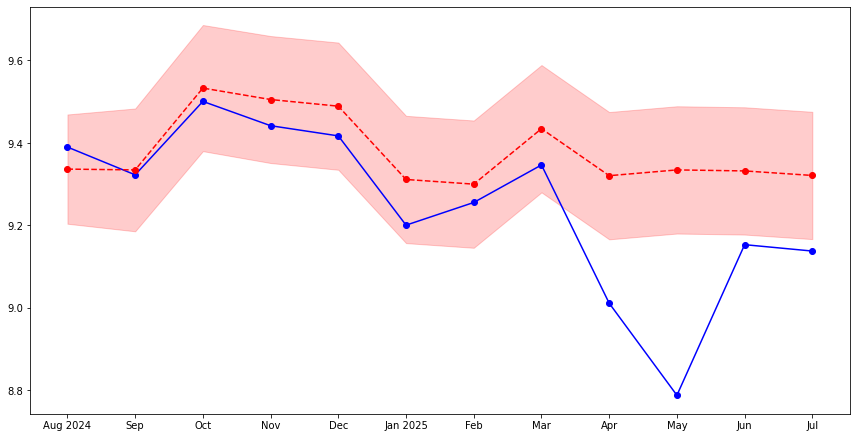}
    \caption{Observed values of the test subsample (blue) and the corresponding mean forecasts with a 95\% prediction interval (red dotted line and shaded area).}
    \label{fig:retail_forecast}
\end{figure}

Figure~\ref{fig:retail_forecast} shows the mean predicted values for the test subsample with 95\% prediction intervals, assuming normally distributed errors. The predictions prior to April 2025 are reasonably accurate, capturing the gradually declining trend and the corresponding oscillations around it. However, from April onward the forecasts deteriorate, primarily due to recent geopolitical tensions and escalating tariff measures on both sides. Therefore, as noted earlier, the forecasts are reliable only when no structural shift occurs in the test sample and when the final trend estimated in the training sample continues to hold in the future. Overall, the STSA model provides interpretable insights into U.S. exports of goods to Mainland China by accurately identifying structural breaks, quantifying regime-specific trends, capturing seasonal effects, characterizing the error structure, and producing reliable short-term forecasts. This example illustrates that, unlike most conventional time series models, the proposed approach not only fits the data and generates forecasts but also uncovers the underlying narrative of the series. By explicitly capturing the influence of external events, it provides a causal interpretation of how external events shape the observed dynamics.

\subsection{Forecasting Performance}

In this section, we evaluate the forecasting performance of the STSA model relative to other statistical approaches using the 48{,}000 monthly series from the M4 Competition \citep{makridakis_etal_2020, godahewa2020m4}. The dataset spans a wide range of domains, including macroeconomic and microeconomic indicators, industry, finance, demographics, and others, with sample sizes ranging from 60 to 2{,}812 observations. The forecast horizon is set to $f = 12$ months. To stabilize the variance, Box–Cox transformation is applied to each series, with the transformation parameter $\lambda \in (-2, 2)$ selected according to the method described by \citet{guerrero1993time}. The parameters of the STSA model are set according to the values recommended before. As the model must be run across many time series, some with very large sample sizes, we implement the parameter $s$ in the modified dynamic programming algorithm, which represents the step size for candidate break points. For example, in the single-break case, the possible locations for the break are $T_1 \in \{l_1, l_1+s, l_1+2s, \dots \}$. We adjust $s$ based on the sample size $T$, setting $s = \lfloor T/200 \rfloor + 1$. This parameter has a negligible effect on the model's accuracy but significantly reduces computation time.

Apart from the previously mentioned models, ARIMA and Prophet, we also consider several other well-established univariate statistical models, including exponential smoothing (ETS; \citealp{hyndman2008forecasting}), Trigonometric Box--Cox ARMA Trend Seasonal (TBATS; \citealp{delivera2011modeling}), and Theta \citep{assimakopoulos2000theta}. These models, except for Prophet, are considered state-of-the-art methods for forecasting of short- to medium-length series, particularly for monthly data from the M4 Competition \citep{godahewa2021monash}. Additionally, we report results for the naive seasonal model, which generates predictions using the last 12 observations of the test sample. The forecasting performance of each model is evaluated using the mean absolute scaled error (MASE) proposed by \citet{hyndman2006another}, defined as
\begin{equation}
\text{MASE} = 
\frac{\frac{1}{f} \sum_{k=1}^{f} |\hat{y}_{T+k} - y_{T+k}|}
{\frac{1}{T-P} \sum_{t=P+1}^{T} |y_t - y_{t-P}|},
\label{eq:mase}
\end{equation}
where $T$ is the number of observations in the training sample, $f$ is the forecast horizon, and $P$ is the seasonal period of the time series; for $k = 1, \dots, f$, $y_{T+k}$ and $\hat{y}_{T+k}$ denote the actual and forecasted values within the test sample, respectively. MASE measures forecast accuracy by scaling the mean absolute error of a model relative to the in-sample mean absolute error from a naive (seasonal random walk) forecast, making it comparable across series of different scales. The Prophet model was implemented using the \texttt{fbprophet} library \citep{prophet}, while the other benchmark models were automated using the \texttt{sktime} library \citep{sktime1, sktime2}. All models were run with their default configurations, although further hyperparameter tuning and domain-specific adjustments could potentially improve their performance.

\begin{figure}[t]
    \centering
    \includegraphics[width=0.9\linewidth]{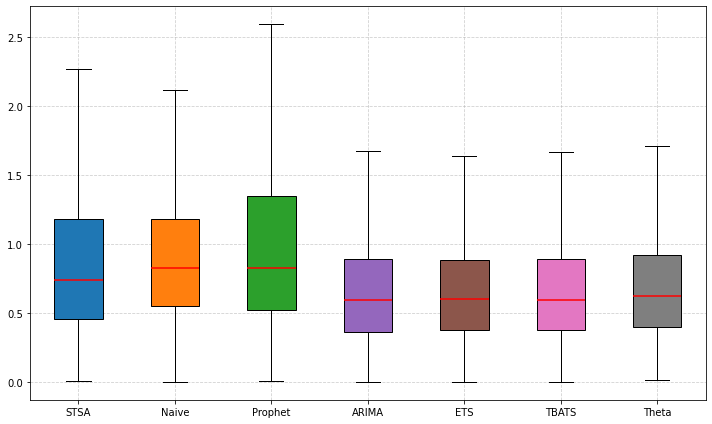}
    \caption{Box-Whisker plot of MASE values for different forecasting models. The boxes represent the interquartile range, the whiskers indicate the range, and the solid red line marks the median. Outliers are not shown.}
    \label{fig:box_whisker}
\end{figure}

Figure~\ref{fig:box_whisker} displays the distribution of MASE values for each forecasting model across all series using a box-whisker plot. Table~\ref{tab:stsa_summary2} presents the corresponding summary statistics, including each model's win and loss rates alongside their average ranks. The win rate is computed as the proportion of series on which a model attains the lowest MASE, while the loss rate is defined analogously as the proportion of series on which it attains the highest MASE. The average rank is computed as the mean positional rank of each model across all series based on its MASE performance. To mitigate the impact of extreme outliers caused by numerical optimization or other model-specific issues, MASE values above the 95th percentile are trimmed to prevent disproportionate distortion of the estimated mean and standard deviation. The best performing model for each summary statistic is highlighted in bold. 

These results show that stochastic trend models—particularly ARIMA, ETS, TBATS, and the Theta model—substantially outperform STSA, Naive, and Prophet. However, STSA still exceeds the performance of the naive seasonal model and, importantly, outperforms Prophet, which employs a similar decomposition framework based on deterministic components with structural breaks. The relatively large standard deviation of STSA, exceeding even that of the Naive model, indicates instability and suggests that its performance varies substantially across series. Its comparatively high loss rate further shows that it is not well suited to many settings. Nevertheless, it achieves a win rate of nearly $15.2\%$, outperforming the Naive, Prophet, ETS, and Theta models, indicating strong performance on a specific subset of series. Overall, although STSA is designed primarily for inference rather than forecasting, it still delivers competitive predictive accuracy, even if it falls short of the more flexible stochastic trend models. 

\begin{table}[t]
\centering
\caption{Summary statistics of MASE values for different forecasting models.}
\begin{tabular}{lcccccccc}
\toprule
 & STSA & Naive & Prophet & ARIMA & ETS & TBATS & Theta \\
\midrule
Mean & 0.818 & 0.851 & 0.972 & \textbf{0.620} & 0.624 & 0.627 & 0.652 \\
Median & 0.742 & 0.826 & 0.826 & \textbf{0.591} & 0.601 & 0.596 & 0.623 \\
Std. Dev. & 0.499 & 0.412 &  0.675 & 0.346 & \textbf{0.332} & 0.337 & 0.345 \\
Win Rate & 0.152 & 0.098 & 0.124 & \textbf{0.193} & 0.135 & 0.156 & 0.142 \\
Loss Rate & 0.226 & 0.283 & 0.316 & 0.052 & \textbf{0.031} & 0.043 & 0.049 \\
Average Rank & 4.377 & 4.915 & 4.760 & \textbf{3.389} & 3.430 & 3.486 & 3.642 \\
\bottomrule
\end{tabular}
\label{tab:stsa_summary2}
\end{table}

\begin{figure}[t]
    \centering
    \begin{subfigure}[b]{0.9\textwidth}
        \includegraphics[width=\textwidth]{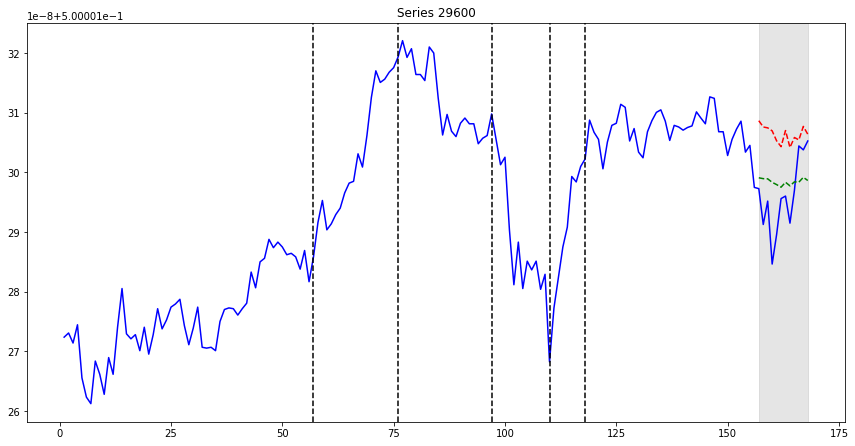}
    \end{subfigure}
    \begin{subfigure}[b]{0.9\textwidth}
        \includegraphics[width=\textwidth]{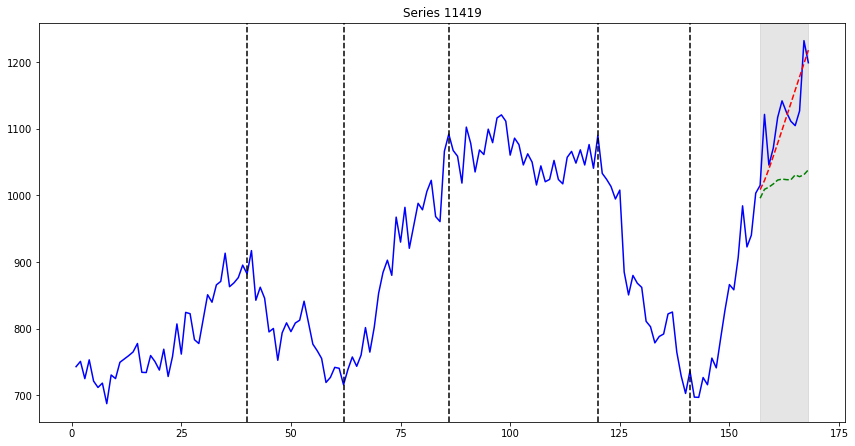}
    \end{subfigure}
    \caption{Scenarios where STSA underperforms and outperforms stochastic trend models. Vertical dotted lines and the red line indicate breaks and STSA predictions, while the green line shows the average forecast of ARIMA, ETS, TBATS, and Theta models.}
    \label{fig:forecast}
\end{figure}

The primary forecasting advantage of stochastic trend models over deterministic approaches such as STSA and Prophet lies in how they handle the most recent observations. Although STSA can accommodate structural changes, it is less responsive to irregular fluctuations near the forecast horizon. Stochastic trend models, by contrast, adapt directly to the latest data, making them more flexible for forecasting. In essence, while STSA infers the underlying trend from recent observations and therefore follows a relatively rigid functional form, stochastic models update their components directly from the latest data, enabling them to capture short-term dynamics that STSA typically treats as transitory shocks rather than meaningful signals. The top panel of Figure~\ref{fig:forecast} illustrates a typical example of this phenomenon, showing that stochastic trend models are more flexible in capturing the most recent movements of the series. 

This limitation of the STSA model can be mitigated by relaxing the parameters responsible for break detection. However, doing so may reduce interpretability by generating an excessive number of detected breaks, complicating the analysis of the series' dynamics. Additionally, it can result in the identification of spurious breaks near the end of the training sample, causing overfitting and misrepresenting the trend of the final regime, which in turn leads to poor forecasts. To enhance forecasting performance, we recommend conducting a visual preliminary analysis to determine appropriate break-detection parameters. Overall, the accuracy of the STSA model depends on the proper identification of structural breaks and the stability of the series near the end of the sample.

In particular, the model performs best when breaks are pronounced and well-separated, allowing precise estimation of both their number and locations. STSA demonstrates a clear advantage when a series exhibits infrequent structural changes, particularly those that alter the long-term slope or direction of the trend. Stochastic trend models typically assume that the underlying data generating process is stable over time, with parameters that evolve smoothly and without abrupt shifts. As a result, when a structural break occurs, these models fail to fully adjust to the sudden change, producing conservative forecasts that often flatten toward the last estimated level rather than reflecting the new trend direction. STSA, on the other hand, explicitly models these shifts through a piecewise linear trend with estimated breaks. The bottom panel of Figure~\ref{fig:forecast} illustrates a case where the STSA model successfully captures the emerging local trend in the final segment of the series, whereas the stochastic trend models fail to adjust to the shift.

The results of this study should not be considered conclusive, as improved parameter tuning could potentially enhance the forecasting performance of the STSA model. Furthermore, the characteristics of the M4 Competition monthly dataset, where most series exhibit slow movements rather than abrupt structural changes, naturally favor stochastic trend models. Nevertheless, the dominance of stochastic models over STSA should not be underestimated, as many real-world series follow patterns that these models capture more flexibly than deterministic trend approaches. At the same time, STSA is better suited for series that experience pronounced structural breaks, such as macroeconomic indicators affected by policy changes, stock prices following major market shocks, or sector-level production or consumption series influenced by sudden economic or technological shifts. In such scenarios, the STSA model generally provides a robust framework that delivers more accurate forecasts by explicitly capturing the post-break trajectory, while stochastic trend models often struggle to adapt.

\section{Conclusion}

This paper introduced the Seasonal-Trend-Stationary ARMA (STSA) model, which posits that nonstationary time series can be expressed as stationary fluctuations around deterministic trend and seasonal components once structural breaks in the trend are incorporated. The model offers an interpretable alternative to prevailing stochastic trend frameworks for time series analysis and forecasting. Methods for estimating break locations were developed using a modified dynamic programming algorithm, and a sequential prediction-interval-based procedure, conceptually similar to cross-validation, was proposed for determining the number of breaks. Simulation results demonstrated that both the number and locations of breaks can be estimated with adequate accuracy. Finally, procedures for specifying and estimating the full model were presented.

Empirical analysis of U.S. goods exports to Mainland China (2006--2025) shows that the STSA model effectively captures major external events, such as the global financial crisis, geopolitical tensions, and COVID-19, through structural breaks that closely align with historical turning points. The estimated components further provide an interpretable description of the series, revealing the trend behavior within each regime, seasonal effects, and the temporal dependency structure. Analysis of forecasting performance on the monthly M4 Competition dataset shows that the proposed model outperforms its conceptually similar counterpart, Prophet, but generally falls short of state-of-the-art stochastic approaches such as ARIMA, ETS, TBATS, and Theta. These models are better suited for forecasting due to their flexible stochastic nature, whereas STSA often underperforms because of its rigid functional form and reliance on user-specified parameters. The main strength of the proposed model appears when the series exhibits pronounced, abrupt structural breaks, where stochastic trend models typically struggle to adjust effectively.

Several limitations remain. The current implementation treats the estimated break dates as known and does not provide confidence intervals, which may be important in some applications. Although the procedure for selecting the number of breaks performs well, it could be improved by accounting for the true distribution of the test statistic and the uncertainty in the pre-estimated parameters of the seasonal and ARMA components. 
Additionally, the parameters of the STSA model related to break detection and model selection could be made data-dependent rather than relying on rule-of-thumb values or visual inspection of the series. This is especially important in large-scale forecasting tasks, where each series possesses its own distinct features.

A more fundamental limitation concerns the assumption that the series are stationary around deterministic components with structural changes. When the errors exhibit strong persistence, this assumption may be violated, and the procedure may fail to produce meaningful parameter estimates. In such cases, it may be more appropriate to assume that the errors contain a unit root. This challenge reflects the well-known circular dependence between testing for structural breaks and evaluation stationarity, since the correct specification of each test depends on prior knowledge of the other. This issue has been extensively studied in the literature, with \citet{perron2009testing} discussing it in the context of break testing and \citet{kejriwal2010sequential} proposing joint estimation procedures to account for this interdependence.

Despite these challenges, the STSA framework remains a theoretically sound and interpretable alternative to stochastic trend models. Unlike many standard methods, it goes beyond data fitting and forecasting by providing parameters that reveal the narrative behind the series’ evolution. By explicitly linking structural changes to regime‐dependent dynamics, STSA reduces reliance on abstract components and offers a transparent account of how observable events shape the series. This interpretability strengthens explanatory power and enhances the clarity and credibility of the resulting forecasts.

\newpage

\bibliography{references}

\end{document}